\documentclass[letterpaper,twocolumn,superscriptaddress,
	    nobibnotes,aps,prb,longbibliography,floatfix]{revtex4-1}
\usepackage[utf8]{inputenc}
\usepackage[T1]{fontenc}
 
\usepackage{amsmath,amssymb,amsfonts}
\usepackage{amstext, mathrsfs, textcomp}
\usepackage{nicefrac}
\usepackage{multirow}

\usepackage{hyperref}
\usepackage{subfigure}
\usepackage{graphicx}
\usepackage{xcolor}
\usepackage[export]{adjustbox}

\usepackage[final]{changes}

\graphicspath{{Figures/}{TOC_ACS/}}

\newcommand{\FIGCAPTIONPREFIX}{}

\newcommand{\TITLE}{Strongly directional scattering from dielectric nanowires}

\begin{document}

%%%%%%%%%%%%%%%%%%%%%%%%%%%%%%%%%%%%%%%%%%%%%%%%%%%%%%%%%%%%%%%%%%%%%%%%%
%% REVTEX
%%%%%%%%%%%%%%%%%%%%%%%%%%%%%%%%%%%%%%%%%%%%%%%%%%%%%%%%%%%%%%%%%%%%%%%%%
\title{\TITLE}
% \subtitle{}

\author{\firstname{Peter R.} \surname{Wiecha}}
\email[e-mail~: ]{peter.wiecha@cemes.fr}
\affiliation{CEMES-CNRS, Universit\'e de Toulouse, CNRS, UPS, Toulouse, France}

\author{\firstname{Aurélien} \surname{Cuche}}
\affiliation{CEMES-CNRS, Universit\'e de Toulouse, CNRS, UPS, Toulouse, France}

\author{\firstname{Arnaud} \surname{Arbouet}}
\affiliation{CEMES-CNRS, Universit\'e de Toulouse, CNRS, UPS, Toulouse, France}

\author{\firstname{Christian} \surname{Girard}}
\affiliation{CEMES-CNRS, Universit\'e de Toulouse, CNRS, UPS, Toulouse, France}

\author{\firstname{Gérard} \surname{Colas des Francs}}  % 2D-GDM
\affiliation{ICB, UMR 6303 CNRS - Universit\'e Bourgogne-Franche Comt\'e, Dijon, France}

\author{Aurélie Lecestre}  % EBL litho
\affiliation{LAAS-CNRS, Universit\'e de Toulouse, CNRS, INP, Toulouse, France}

\author{Guilhem Larrieu}  % EBL litho
\affiliation{LAAS-CNRS, Universit\'e de Toulouse, CNRS, INP, Toulouse, France}

\author{Frank Fournel}  % SOQ Substrate
\affiliation{CEA-LETI/MINATEC, CEA, Grenoble, France}

\author{Vincent Larrey}  % SOQ Substrate
\affiliation{CEA-LETI/MINATEC, CEA, Grenoble, France}

\author{Thierry Baron}  % VLS SiNWs
\affiliation{CNRS, LTM, Universit\'e Grenoble Alpes, Grenoble, France}

\author{\firstname{Vincent} \surname{Paillard}}
\email[e-mail~: ]{vincent.paillard@cemes.fr}
\affiliation{CEMES-CNRS, Universit\'e de Toulouse, CNRS, UPS, Toulouse, France}
%%%%%%%%%%%%%%%%%%%%%%%%%%%%%%%%%%%%%%%%%%%%%%%%%%%%%%%%%%%%%%%%%%%%%%%%%
%%%%%%%%%%%%%%%%%%%%%%%%%%%%%%%%%%%%%%%%%%%%%%%%%%%%%%%%%%%%%%%%%%%%%%%%%

%% ACHEMSO Graphical TOC entry
% \begin{tocentry}
% \includegraphics{"Graphical_TOC"}
% \end{tocentry}

\begin{abstract}
It has been experimentally demonstrated only recently that a simultaneous excitation of interfering electric and magnetic resonances can lead to uni-directional scattering of visible light in zero-dimensional dielectric nanoparticles.
We show both theoretically and experimentally, that strongly anisotropic scattering also occurs in individual dielectric nanowires. 
The effect occurs even under either pure transverse electric or pure transverse magnetic polarized normal illumination.
This allows for instance to toggle the scattering direction by a simple rotation of the incident polarization.
Finally, we demonstrate that directional scattering is not limited to cylindrical cross-sections, but can be further tailored by varying the shape of the nanowires.
\end{abstract}

\maketitle
%% --------------------------------------------------------------
%% Main Text
%% --------------------------------------------------------------

%% ---------------------------------------------- SECTION: Introduction

The search for ways to control light at subwavelength dimensions has increasingly attracted the interest of researchers for about the last two decades.
Due to their strong polarizability and tunable plasmon resonances, metallic nanostructures are particularly suitable for the nanoscale manipulation of light -- especially at visible frequencies.\cite{muhlschlegel_resonant_2005}
However, such plasmonic structures suffer from certain drawbacks like strong dissipation associated to the large imaginary part of the dielectric function in metals.

Recently, dielectric nanostructures from high-index materials have proven to offer a promising alternative platform with far lower losses. \cite{kuznetsov_optically_2016}
Like in plasmonics, it is possible to tune optical resonances from the near ultra-violet to the near infrared, yet with almost no dissipative losses.
At these resonances, which can be of both electric or magnetic nature, strong local field enhancements \cite{bakker_magnetic_2015} and intense scattering \cite{kuznetsov_magnetic_2012} occur, tunable via the material and the geometry of the nanostructure. 
Prominent dielectric materials are, among others, silicon, germanium or III-V compound semiconductors with indirect band-gap. \cite{kallel_tunable_2012, albella_electric_2014} 
Conventional geometries include spherical nanoparticles \cite{kuznetsov_magnetic_2012} or nanowires (NWs), \cite{cao_tuning_2010, ee_shape-dependent_2015} but also more complex dielectric nanostructures \cite{wiecha_evolutionary_2017}. 
Dielectric optical antennas are promising candidates for applications in field-enhanced spectroscopy, \cite{gerard_strong_2008, wells_silicon_2012, regmi_all-dielectric_2016, cambiasso_bridging_2017} imaging, \cite{kallel_photoluminescence_2013} to enhance and control nonlinear effects \cite{shcherbakov_enhanced_2014, wiecha_enhanced_2015, wiecha_origin_2016} or to increase the efficiency in photovoltaics \cite{brongersma_light_2014}.

A peculiarity of dielectric particles is the possibility to simultaneously obtain a strong electric and magnetic response using very simple geometries. \cite{kuznetsov_magnetic_2012, bakker_magnetic_2015, mirzaei_electric_2015, valuckas_direct_2017}
Recently, it has been independently shown by two research groups,\cite{person_demonstration_2013, fu_directional_2013} that exclusive forward (FW) or backward (BW) scattering, predicted by Kerker \textit{et al.} in 1983 for hypothetical magneto-dielectric particles, \cite{kerker_electromagnetic_1983} can be realized in the visible spectral range using dielectric nanoparticles. 
Kerker \textit{et al.} described two possible configurations, called the \textit{Kerker conditions}. 
At the first Kerker condition zero backward scattering occurs for equal electric permittivity and magnetic permeability (\(\epsilon_r = \mu_r\)).
The second Kerker condition predicts zero forward scattering in small spherical particles when the first order magnetic and electric Mie coefficients are of equal absolute value and with opposite sign (\(a_1 = -b_1\)).
In contrast to particularly designed metamaterials,\cite{pendry_controlling_2006} the magnetic permeability \(\mu_r\) is unitary in dielectric nanoparticles. 
Nevertheless, simultaneously occurring electric and magnetic resonances can \textit{de-facto} fulfill the first Kerker condition. \added{\cite{gomez-medina_electric_2011,nieto-vesperinas_angle-suppressed_2011}}
While the Kerker conditions were originally derived for spherical particles, it has been shown that the conditions are a result of a cylindrical symmetry and therefore can be generalized accordingly. \cite{zambrana-puyalto_duality_2013}

%%-------------------------- FIGURE: cyl. Mie Field enhancement ---
\begin{figure}[t!]
\centering	
\includegraphics[page=1]{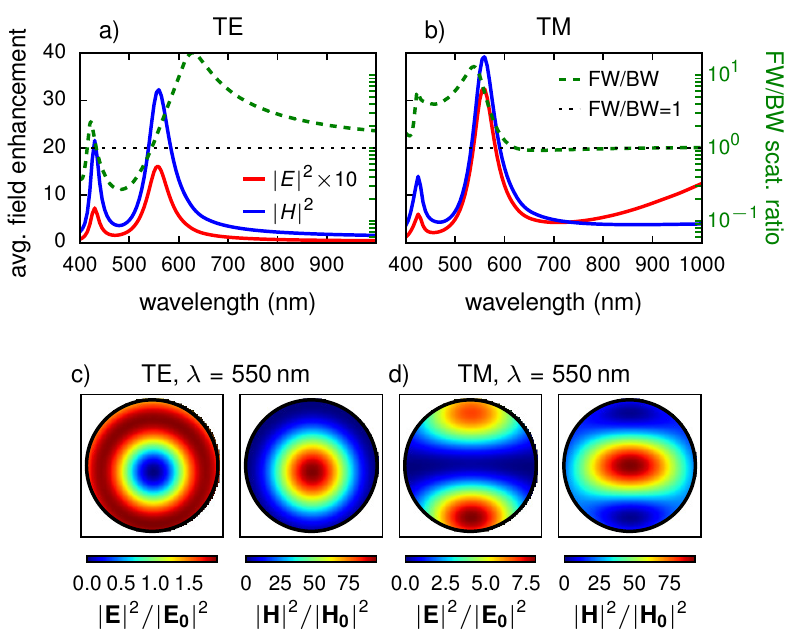}
\caption{\FIGCAPTIONPREFIX
Average electric (red lines) and magnetic (blue lines) field intensity enhancement inside a silicon nanowire of diameter \(D=100\,\)nm as function of wavelength for a) a TE and b) a TM polarized incident plane wave, calculated using Mie theory. 
Normalized to the illumination field intensity.
For comparison, the FW/BW scattering ratio in the far-field is shown as dashed green line (right axis ticks).
c) and d) show the internal field intensity distributions at \(\lambda=550\,\)nm for TE and TM polarization, respectively (left subplot: electric, right: magnetic field). \(D=100\,\)nm, plane wave incident from the top.
}
\label{fig:fig1}
\end{figure}
%%-----------------------------------------------------------------

Recent publications have confirmed the possibility to obtain optimum FW scattering also from elongated structures such as nanopillars, \cite{person_demonstration_2013} spheroids \cite{lukyanchuk_optimum_2015} or even from cuboidal dielectric particles \cite{ee_shape-dependent_2015}.
Other dielectric particles on which directional scattering was investigated include  nanodiscs for FW/BW directional metasurfaces, \cite{staude_tailoring_2013, decker_resonant_2016} patch antennas, \cite{yang_controlling_2015} V-shaped structures for multi-directional color routing, \cite{li_all-dielectric_2016} or asymmetric hollow nanodiscs for bi-anisotropic scattering \cite{alaee_all-dielectric_2015}, all of them based on the interplay between electric and magnetic modes.
% Dielectric metasurface: Disorder to couple Mie/Guided modes and obtain omni-directional scattering \cite{lin_frequency-selective_2016}.
%
Control over the directional scattering can also be obtained using arrangements of nanoparticles like dimers \cite{albella_switchable_2015, campione_tailoring_2015, yan_directional_2015} or via hybrid metal/dielectric nano-structures \cite{guo_multipolar_2016}.
Even directional shaping of nonlinear emission has been demonstrated: 
The radiation pattern of the third harmonic generation from silicon dimers can be controlled via the geometry of the structure. \cite{wang_shaping_2017}
In a plasmonic-dielectric hybrid structure, the interplay of electric and magnetic resonances in dielectric TiO\(_2\) spheres was used to impose an uni-directional radiation pattern on the second harmonic generation from a plasmonic driver element.\cite{xiong_compact_2016}

Compared to zero-dimensional (0D) particles \added{of deeply sub-wavelength size in all directions (like nano-spheres)}, \added{one-dimensional (1D)} nanowires have a strongly polarization dependent optical response, which offers an additional degree of freedom, supplementary to the choice of material and modifications of the \added{nano-}structure's geometry. 
Despite this opportunity, research on directional scattering from \deleted{one-dimensional (1D)} dielectric nanowires is still scarce.
\replaced{As for scattering, it has been shown theoretically that}{As for scattering from single nanowires, it has been shown theoretically that} multi-layer dielectric \cite{mirzaei_all-dielectric_2015} or plasmonic-dielectric core-shell \cite{liu_invisible_2015} nanowires can be rendered invisible by tailoring destructive interference between \replaced{different order modes}{electric or magnetic} \added{or electric and toroidal dipole} modes\added{, respectively}. 
Such entities might be useful for cloaking applications, for instance to create ``invisible'' electric contacts or circuits. \cite{fan_invisible_2012}
\added{Dense arrays of vertical dielectric NWs on the other hand, possess a strong light extinction due to multiple scattering inside the nanowire assembly, which can be used as an anti-reflection coating for photovoltaics.\cite{grzela_polarization-dependent_2012, muskens_design_2008} 
At the level of an individual nanowire, angle-dependent absorption can be tailored via the incident angle through the interplay between Mie-type and guided modes.\cite{abujetas_unraveling_2015}
Also, the directionality of photoluminescence from individual III-V nanowires can be tailored through the supported Mie and guided modes \cite{grzela_nanowire_2012, paniagua-dominguez_enhanced_2013, brenny_directional_2016}.}
%
%
% \added{For a broader overview on directional scattering and emission from individual nanoparticles, we refer to Ref.~\onlinecite{wei_control_2014}.}

\added{In contrast to these former works on complex geometries or effects, in this paper we study the directional scattering of light from individual normally illuminated dielectric nanowires.}
Using Mie theory we analyze in a first step the interplay of different \added{order Mie} modes, leading to the occurrence of directional scattering.
We then compare experimental results from \added{individual, single crystal} silicon nanowires (SiNWs) of cylindrical and rectangular cross-sections. 
In both cases spectral zones of strongly anisotropic FW/BW scattering ratios can be identified and we find that asymmetric wire geometries allow even further tailoring of directional scattering \added{and offer the incident angle of the illumination as a supplementary free parameter}. 
We confront our experimental results with simulations using the Green dyadic method (GDM), yielding a very good agreement.

%% ---------------------------------------------- SECTION: Directional scattering Mie 
\section{Directional scattering from cylindrical nanowires}

Mie theory can be applied to infinitely long cylinders by expanding the fields in vector cylindrical harmonics. 
The Mie scattering coefficients \(a_i\) and \(b_i\) of the expansion can be regarded as weights for corresponding electric and magnetic multipole moments, representing the response of the wire to an external illumination.
Under normal incidence, the Mie S-matrix, connecting the incident (\(\mathbf{E}_i\)) and the asymptotic, scattered field (\(\mathbf{E}_s\)) writes\cite{bohren_absorption_1998}
\begin{equation}\label{eq:scattering_Smatrix}
 \begin{bmatrix}
  E_{s, \text{TM}} \\
  E_{s, \text{TE}}
 \end{bmatrix}
  = 
e^{\mathrm{i} 3\pi/4} \, \sqrt{\dfrac{2}{\pi k R}} \, e^{\mathrm{i} k R}
 \begin{bmatrix}
  T_1 & 0    \\
  0   & T_2
 \end{bmatrix}
 \begin{bmatrix}
  E_{i, \text{TM}} \\
  E_{i, \text{TE}}
 \end{bmatrix}
\end{equation}
with the wavenumber \(k=2\pi/\lambda\), the distance \(R\) to the cylinder axis and
\begin{equation}\label{eq:Smatrix_T1_T2}
 \begin{aligned}
  T_1 & = b_0 + 2 \sum\limits_{n=1}^{\infty} b_n \cos(n \varphi) \\
  T_2 & = a_0 + 2 \sum\limits_{n=1}^{\infty} a_n \cos(n \varphi). 
 \end{aligned}
\end{equation}
\(\varphi\) is the scattering angle with respect to the incident wave vector and \(\varphi = 0\) corresponds to the forward scattering direction.

As can be seen from Eq.~\eqref{eq:scattering_Smatrix}, under normal incidence the transverse magnetic (TM) and transverse electric (TE) polarized components of the scattered fields are proportional to the S-matrix components \(T_1\), respectively \(T_2\).
Therefore, according to Eqs.~\eqref{eq:Smatrix_T1_T2} scattering from a TE polarized normally incident plane wave (\(\mathbf{E} \perp\) NW axis) is only due to the ``electric'' multipole contributions~\(a_i\).
On the other hand, a TM polarized illumination (\(\mathbf{E} \parallel\) NW axis) induces scattering exclusively via the ``magnetic'' Mie terms~\(b_i\).
The expressions ``electric'' and ``magnetic'' refer to the fact that in the TE and TM case, the magnetic, respectively electric field components are zero in the scattering plane.

%%--------------------------------------- FIGURE: Exp. Setup ---
\begin{figure}[t!]
\centering
\includegraphics[width=\columnwidth,page=1]{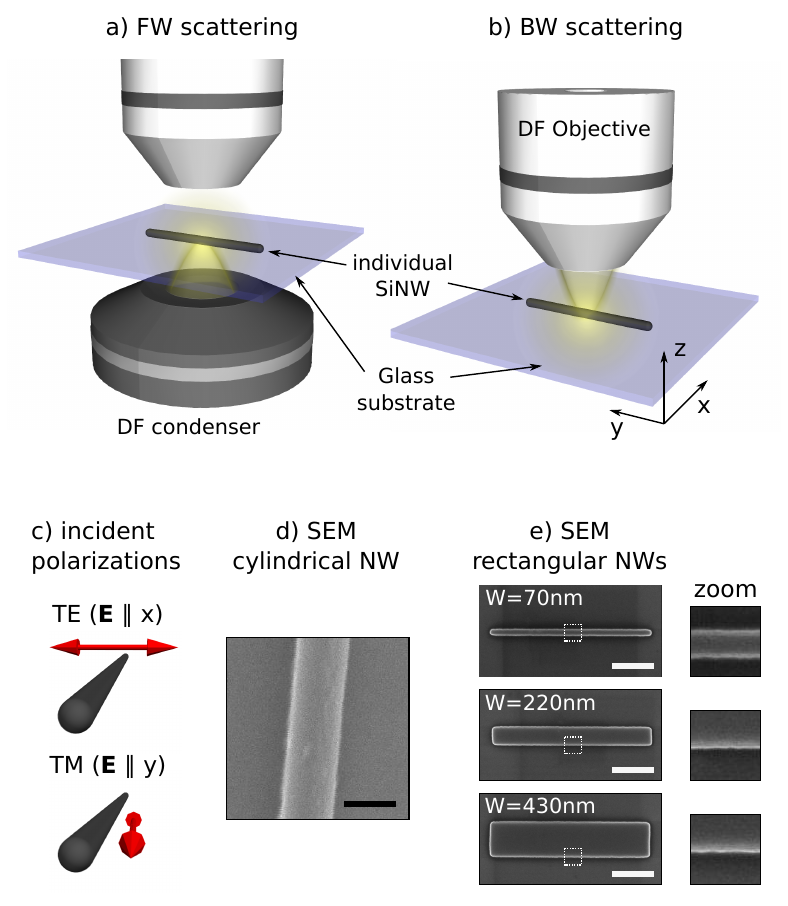}
\caption{\FIGCAPTIONPREFIX
Illustration of the experimental setup for (a) forward scattering and (b) backward scattering measurements. 
(c) sketch of the incident polarization configurations. The electric field is orientated either perpendicular (TE) or parallel (TM) to the NW axis.
(d) SEM image of a cylindrical SiNW, deposited on a silicon substrate. The NWs for the spectroscopy experiments were deposited on a glass substrate, which could not be imaged in SEM due to charging. Scale bar is \(100\,\)nm.
(e) SEM images of rectangular SiNWs. SEM images are obtained from a second sample on silicon on insulator (SOI) substrate. The spectroscopy sample is fabricated on transparent, but insulating SOQ substrate. 
Scale bars are \(500\,\)nm. 
Zoomed insets on the right are \(200 \times 200\,\)nm\(^2\) (white dashed squares indicate zoom area).
}
\label{fig:fig2}
\end{figure}
%%---------------------------------------------------------------

%%------------------------------ FIGURE: cyl-SiNW Spectra / Fields ---
\begin{figure*}[t!]
\centering	
\includegraphics[page=1]{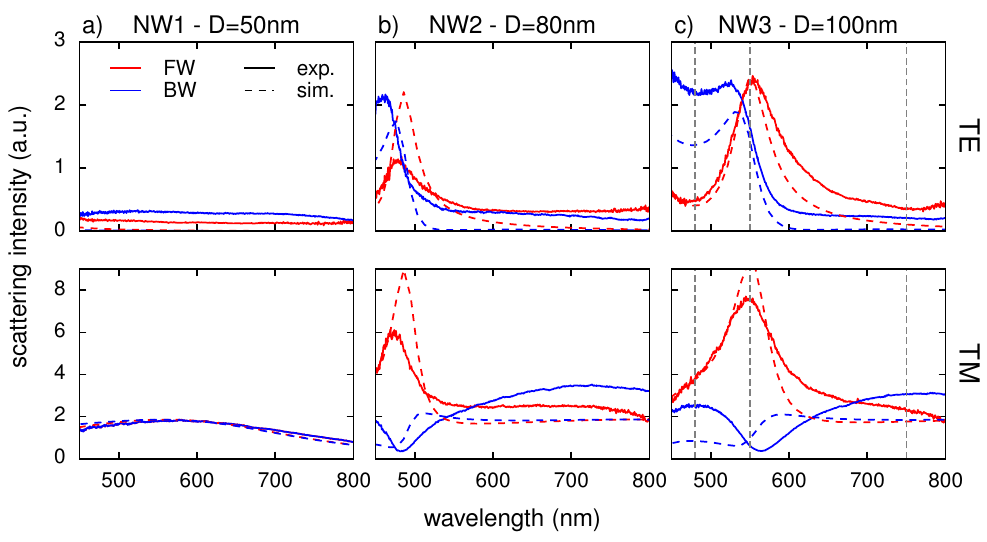}
\hspace{0.5cm}
\includegraphics[page=1]{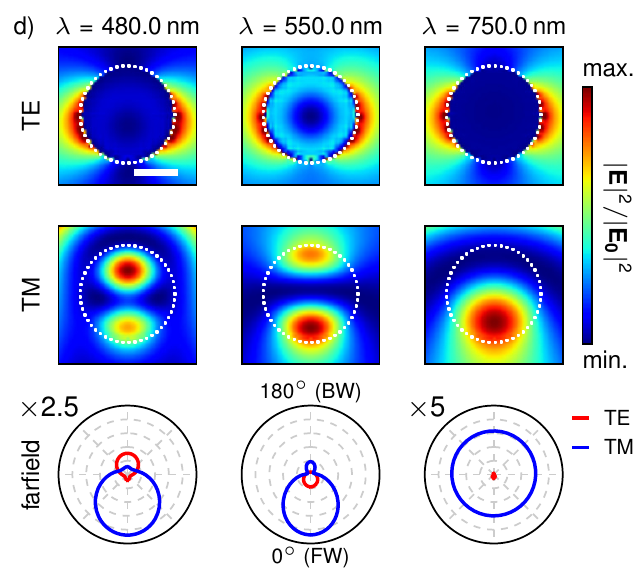} 
\caption{\FIGCAPTIONPREFIX
Experimental (solid lines) and simulated (dashed lines) FW/BW scattering spectra (red/blue) from cylindrical NWs of different diameters. (a) \(D\approx 50\,\)nm, (b) \(D\approx 80\,\)nm and (c) \(D\approx 100\,\)nm. Diameters are estimates by comparison to simulations. 
Top row and bottom row show the case of TE and TM polarized incident plane waves, respectively.
(d) calculated nearfield distributions (top row: TE, center row: TM incidence) and farfield patterns (bottom row; TE/TM: red/blue) for a nanowire of \(D = 100\,\)nm diameter at selected wavelengths, indicated by dashed vertical lines in (c). Plane wave incident from the top. Scale bar is \(50\,\)nm.
}
\label{fig:fig3}
\end{figure*}
%%------------------------------------------------------------------

Let us now consider the case of sufficiently small nanowires, where only the first two orders of the Mie expansion contribute significantly to scattering.
For SiNWs in the visible spectral range, this assumption is a good approximation for diameters up to at least \(D=100\,\)nm ({\color{blue}see also supporting informations (SI), Figs.~S1-S2}).
By setting \(\varphi=0\) or \(\varphi=\pi\) in equations~\eqref{eq:Smatrix_T1_T2} we obtain the conditions for zero scattering in forward (FW), respectively backward (BW) direction. 
For TM polarized illumination we find:
\begin{equation}\label{eq:pure_FW_BW_conditions_TM}
 \begin{aligned}
    T_1\Big|_{\varphi=0} & \approx b_0 + 2 b_1 = 0 \quad\quad \text{for pure BW scat.}\\
    T_1\Big|_{\varphi=\pi} & \approx b_0 - 2 b_1 = 0 \quad\quad \text{for pure FW scat.}
 \end{aligned}
\end{equation}
The same conditions hold for \(T_2\) and \(a_{0,1}\) in the TE configuration.
Hence if the conditions in Eqs.~\eqref{eq:pure_FW_BW_conditions_TM} are met, it is possible to obtain uni-directional scattering from 1D dielectric nanowires, which is the result of interference between the simultaneously excited first two multipole contributions.
In figure~\ref{fig:fig1}(a-b), the FW/BW scattering ratio is shown (dashed green lines) for a SiNW of diameter \(D=100\,\)nm for (a)~TE and (b)~TM polarized excitation, revealing zones of strongly directional scattering.
For the dispersion of silicon we use tabulated data from Ref.~\onlinecite{palik_silicon_1997} throughout this paper.
In the {\color{blue}SI Figs~S3-S4} we show the first two order Mie coefficients for SiNWs as function of the wavelength and the nanowire diameter and find that zones where the conditions in Eqs.~\eqref{eq:pure_FW_BW_conditions_TM} are approached do exist at several wavelength / diameter combinations for normally illuminated cylindrical silicon nanowires. 

\added{We want to note that a similar derivation has been performed for core / shell metal / dielectric nanowires, where the plasmonic core was introduced in the NW to shift the electric and magnetic dipolar modes to spectrally overlap, leading to directional scattering effects. \cite{liu_scattering_2013}
In contrast to this former theoretical work, we study interference between different order multipole contributions in \textit{homogeneous} dielectric nanowires, leading to strongly directional scattering even in such very simple systems.
}

%%-------------------------------- FIGURE: Rectwires spectra 2D-maps ---
\begin{figure*}[tb]
\centering
\includegraphics[page=1]{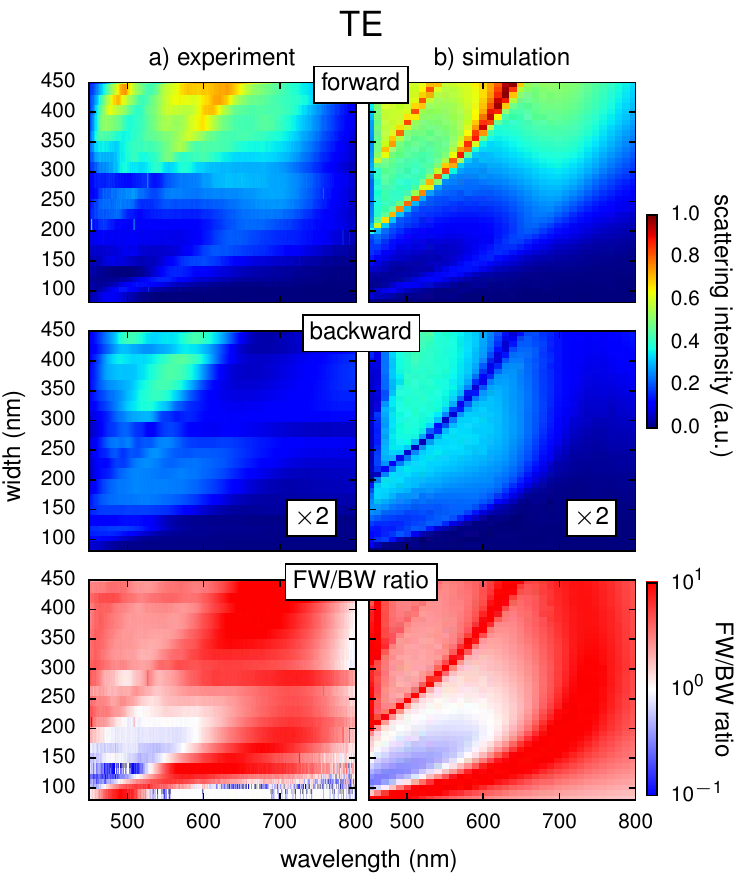} 
\hspace{1cm}
\includegraphics[page=1]{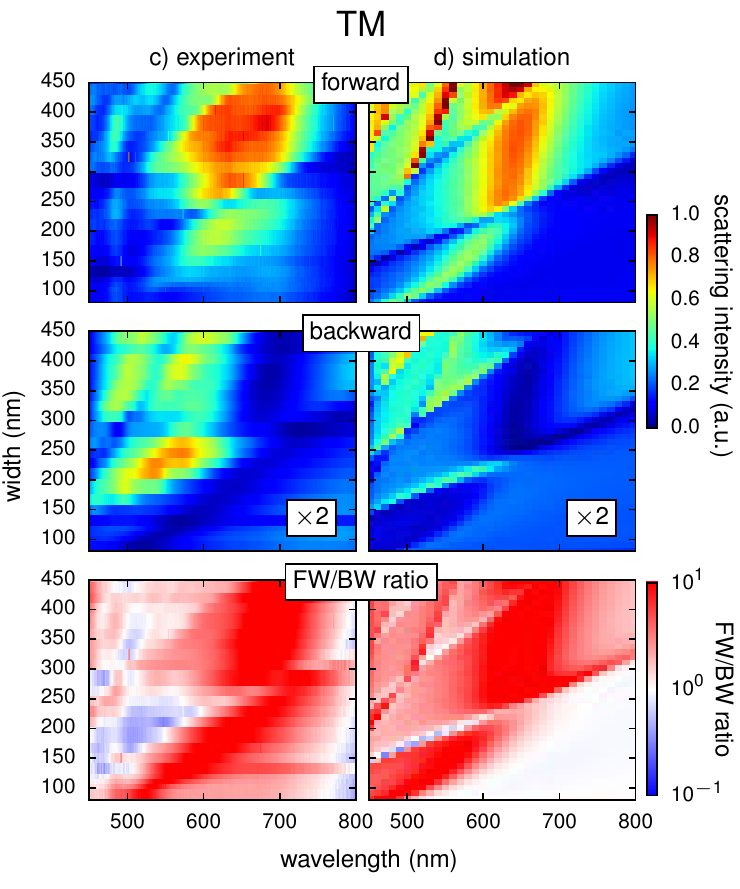} 
\caption{\FIGCAPTIONPREFIX
Forward (top row) and backward (center row, multiplied by \(\times 2\) for better contrast) scattering intensity from rectangular silicon nanowires as function of wavelength and NW width. 
Data is normalized to the global maximum intensity in FW and BW spectra.
FW/BW ratios are shown in the bottom row on a logarithmic color scale.
(a-b) TE, (c-d) TM polarized illumination. 
(a) and (c) show measured dark field spectra, (b) and (d) corresponding GDM simulations.
Fixed NW height and length (\(H=90\,\)nm, \(L=7\,\)\textmu m). 
}
\label{fig:fig4}
\end{figure*}
%%------------------------------------------------------------------

It is possible to expand the electromagnetic fields inside the cylinder in the same way as the scattered fields of Eq.~\eqref{eq:scattering_Smatrix}. 
For details, we refer to the textbook of Bohren and Huffmann (chapters~4 and~8.4).\cite{bohren_absorption_1998}
In figure~\ref{fig:fig1} the average internal field intensity enhancement is shown for a SiNW with diameter \(D=100\,\)nm with a TE~(a) and TM~(b) polarized incident plane wave, respectively.
Interestingly, at the resonant wavelengths we observe not only high electric field intensities (red lines) but also a very strong enhancement of the magnetic field (blue lines), regardless of the incident polarization orientation. 
For both polarizations, the magnetic field increases even significantly stronger compared to the electric field intensity.
We remark that this observation is in agreement with recent results from dielectric cylinders in the GHz regime.\cite{kapitanova_giant_2017}
We conclude that a simultaneous excitation of strong electric and magnetic fields occurs in dielectric, non-magnetic (\textit{i.e.} \(\mu_r = 1\)) nanowires, even under pure TE or TM polarized illumination and normal incidence.
Hence, in analogy to the findings of Kerker \textit{et al.} for the case of spherical dielectric nanoparticles, the observed directionality (dashed green lines in Fig.~\ref{fig:fig1}) can be interpreted as a result of the interference between \textit{``effective'' electric and magnetic modes}.
In particular, no directional scattering is obtained at the non-degenerate, fundamental TM\(_{01}\) mode, where only the internal \textit{electric} field shows a resonant enhancement while the \textit{magnetic} field intensity follows a flat line beyond \(\lambda \gtrsim 700\,\)nm (see Fig.~\ref{fig:fig1}b).
For illustration the electric and magnetic field intensity patterns inside the NW cross section are shown for an incident wavelength \(\lambda=550\,\)nm in Fig.~\ref{fig:fig1}(c-d).

%%------------------------------ FIGURE: RectNW selected spectra/fields ---
\begin{figure*}[tb]
\centering
\includegraphics[page=1]{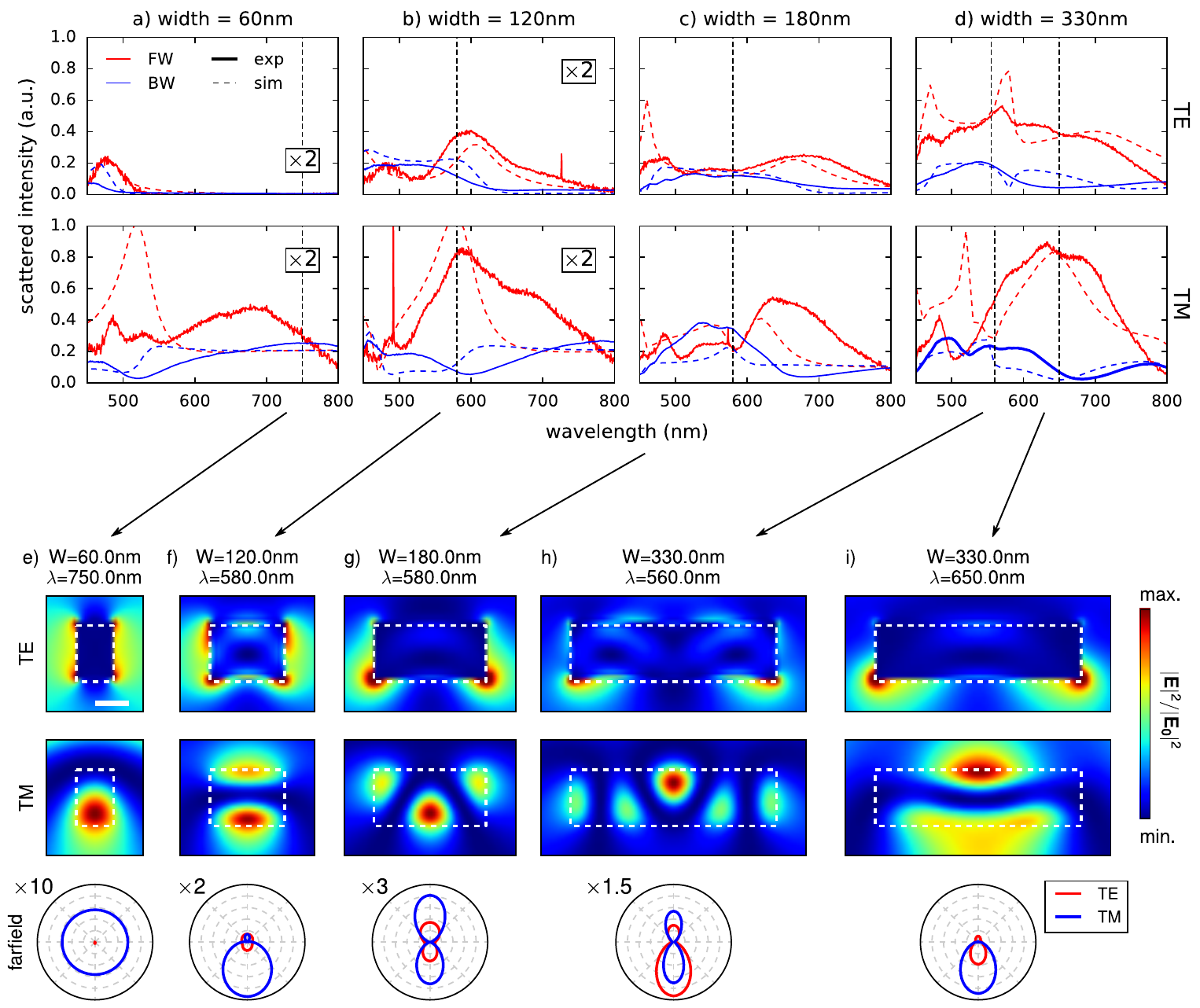}
\caption{\FIGCAPTIONPREFIX
(a-d) selected scattering spectra from rectangular silicon nanowires. Experimental FW (red) and BW (blue) scattered intensity (solid lines) is compared to GDM simulations (dashed lines) for TE (top row) and TM (bottom row) polarized plane wave illumination. 
NW height and length are fixed (\(H=90\,\)nm, \(L=7\,\)\textmu m), widths \(W\) are (a)~\(60\,\)nm, (b)~\(120\,\)nm, (c)~\(180\,\)nm and (d)~\(330\,\)nm, referring to the simulation parameters. 
Intensities in (a-b) are multiplied by \(\times 2\) for better visibility.
Same data as in figure~\ref{fig:fig4}.
Panels (e-i) show selected electric field intensities in the cross section of rectangular SiNWs for (e) \(W=60\,\)nm, \(\lambda=750\,\)nm; (f) \(W=120\,\)nm, \(\lambda=580\,\)nm; (g) \(W=180\,\)nm, \(\lambda=580\,\)nm; (h) \(W=330\,\)nm, \(\lambda=560\,\)nm and (i) \(W=330\,\)nm, \(\lambda=650\,\)nm.
Plane wave illumination from the top, polarized perpendicularly (TE, top row) or along the NW axis (TM, bottom row). 
Farfield radiation patterns are shown in the bottom row (TE: red, TM: blue), adjusted by a variable scaling factor for better visibility (factor indicated on the upper left of each plot).
Scale bar in (e) is \(50\,\)nm, same scale for all field plots.
Vertical dashed lines in (a-d) indicate the spectral positions of the field-plots in (e-i).
}
\label{fig:fig5}
\end{figure*}
%%---------------------------------------------------------------

%% ---------------------------------------------- SECTION: Cylindrical Wires
\section{Spectroscopy on cylindrical silicon nanowires}

In order to compare the forward and backward scattering of a normally incident plane wave on silicon nanowires (SiNWs) we perform standard darkfield microscopy either in reflection (backward scattering, ``BW'') or in transmission geometry (forward scattering, ``FW''). 
The measurement geometry is schematically shown in Fig.~\ref{fig:fig2}(a) and (b) for FW and BW scattering, respectively. 
For details on the measurement technique, see Methods.
We compare our experimental results to 2D~simulations (in \(XZ\)-plane, assuming infinitely long structures along \(Y\)) by the Green dyadic method.
Likewise, the GDM simulation technique is explained in the methods section.
In the {\color{blue}SI Figs.~S5-S10}, the accuracy of the method is verified by comparing GDM simulations to Mie theory and FDTD simulations.

We measure the scattering from cylindrical SiNWs of diameters between \(D\approx 50\,\)nm and \(D\approx 100\,\)nm, epitaxially grown by the vapor-liquid-solid (VLS) technique and drop-coated on a transparent glass substrate.
The NW size-dispersion was obtained via scanning electron microscopy (SEM) from a second sample consisting of the same SiNWs, drop-coated on a silicon substrate (see Fig.~\ref{fig:fig2}d). 
For details, see the methods section.
Results of the scattering experiments are shown in Fig.~\ref{fig:fig3}(a-c) for TE (top row) and TM (bottom row) polarized illumination. 
The given NW diameters are estimates, obtained by comparison with simulations. 

The very weak (TE), respectively omnidirectional (TM) scattering in case of the the smallest nanowire (Fig.~\ref{fig:fig3}a, NW1: \(D\approx 50\,\)nm) is in perfect agreement with the GDM simulations and confirms the assumption of a purely dipolar response if only the non-degenerate TM\(_{01}\) mode is excited (see also field plot for \(\lambda=750\,\)nm in figure~\ref{fig:fig3}d on the right).
As expected, with increasing NW diameter (figure~\ref{fig:fig3}b-c), we observe a simultaneous red-shift of the FW and BW scattering peaks.
For the largest SiNWs (Fig.~\ref{fig:fig3}c, \(D\approx 100\,\)nm), in the case of TE polarized illumination mainly FW scattering occurs in a limited spectral range (\(550\,\text{nm} \lesssim \lambda \lesssim 700\,\)nm), before BW scattering takes over at shorter wavelengths (\(\lambda \lesssim 550\,\)nm).
Note that around \(480\,\)nm, we obtain the possibility to invert the main scattering direction by simply flipping the polarization from TM to TE (see Fig.~\ref{fig:fig3}d, bottom left or also {\color{blue}SI, Fig.~S5}). 
By changing the NW diameter this spectral zone can be also tuned to other wavelengths ({\color{blue}see SI, Figs.~S6-S7}).
At longer wavelengths (\(\lambda\gtrsim 700\,\)nm), only the TM\(_{01}\) mode exists, leading to very weak overall scattering under TE incidence and to the above mentioned, omnidirectional radiation pattern in the TM geometry.
Interestingly, while TE excitation can induce uni-directional BW scattering, under TM polarization BW scattering is generally very weak and mainly FW scattering occurs as soon as higher order contributions are excited together with the TM\(_{01}\) mode.

In summary, we note that although we do observe BW scattering (mainly under TE polarization) it mostly remains weak compared to the FW scattered light. 
This has been also observed in the case of 0D-particles\cite{fu_directional_2013, yan_directional_2015} and can be explained with the finding that the second Kerker's condition (describing BW scattering) \replaced{is contradicting the optical theorem and thus cannot be exactly fulfilled. \cite{alu_how_2010}
Furthermore it is difficult to even closely meet the second condition}{is rather difficult to be met}.\cite{zambrana-puyalto_duality_2013} 
In the case of TM polarized illumination, the FW/BW scattering ratio is even almost exclusively \(\gtrsim 1\) ({\color{blue}see also SI, Figs.~S5-S7} for more simulations on cylindrical NWs).

%%------------------------------ FIGURE: RectNW oblique incidence ---
\begin{figure*}[tb]
\centering
\includegraphics[page=1]{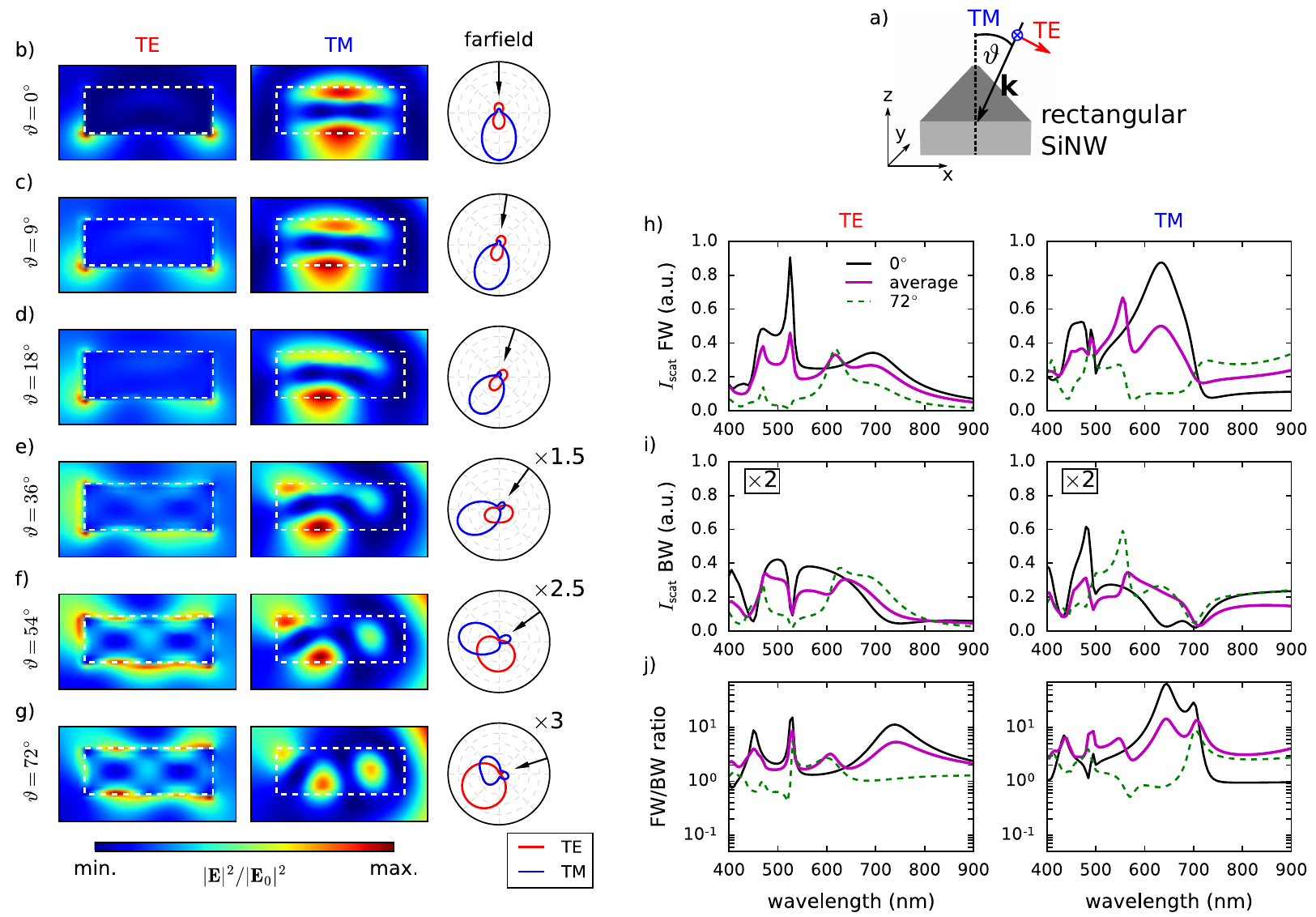}
\caption{\FIGCAPTIONPREFIX
\added{(a) sketch illustrating the oblique incident illumination. 
(b-g) Electric field intensities (individually normalized) in the cross section of rectangular SiNWs of width \(W=250\,\)nm (height \(H=90\,\)nm) for increasing incident angles from normal incidence in (b) to \(\vartheta=72^{\circ}\) in (g). 
The NW cross section is indicated by dashed white lines.
Left: TE, center: TM polarized plane wave of wavelength \(\lambda_0=600\,\)nm. Right: farfield pattern for TE (red) and TM (blue). In the farfield panels, the corresponding angle of incidence is depicted by a black arrow.
(h-j) show corresponding farfield spectra for TE (left) and TM (right) polarization. 
(h) forward scattering (towards lower hemisphere), (i) backward scattering (to upper hemisphere, multiplied by~\(2\)) and (j) FW/BW scattering ratio for normal incidence (black lines), averaged incident angles from~\(0^{\circ}\) to~\(72^{\circ}\) (magenta lines) and \(72^{\circ}\) (green dashed lines).
\(72^{\circ}\) corresponds to the maximum angle of the darkfield illumination in our experiments.}
}
\label{fig:fig6}
\end{figure*}
%%---------------------------------------------------------------

%% ---------------------------------------------- SECTION: Rectangular Wires
\section{Asymmetric, Rectangular Silicon Nanowires}

In a second step we analyze what happens if the cylindrical symmetry of the nanowire cross section is broken. 
We therefore fabricate SiNWs of rectangular section by electron beam lithography (EBL) and subsequent dry-etching on a silicon-on-quartz (SOQ) substrate. 
For details on the fabrication process, see Methods. 
We want to point out a great advantage of our top-down approach on SOQ: The possibility to create \deleted{mono-crystalline} silicon nanostructures of arbitrary shape by EBL on a transparent substrate \added{from single crystalline silicon.
Using crystalline material rather than polycrystalline or amorphous silicon guarantees that the high refractive index of silicon is obtained and facilitates furthermore the modeling, since accurate tabulated data for the dispersion is available in literature.\cite{palik_silicon_1997}}
This gives us the opportunity for a systematic analysis of the scattering from rectangular SiNWs of variable width. 
The height of the structures however remains constant, defined by the thickness of the silicon layer on the SOQ substrate (\(H=90\,\)nm in our case).
We choose SiNWs with length long compared to the focal spot of the illuminating optics (all data acquired on NWs with \(L=7\,\)\textmu m) in order to obtain a purely Mie-like response.\cite{traviss_antenna_2015}
SEM images of selected rectangular SiNWs are shown in figure~\ref{fig:fig2}e. 
The images are obtained from a second sample, fabricated with the exact same process parameters but on a non-insulating silicon-on-insulator (SOI) substrate.
For the purpose of better illustration NWs of \(L=2\,\)\textmu m length are shown. 
We point out that all NWs have excellent surface properties, low roughness and steep flanks. 

The results of our systematic FW/BW scattering measurements are shown in figure~\ref{fig:fig4}. 
Spectra for TE polarized illumination are shown in (a-b), TM spectra in (c-d).
The comparison of experiments (~(a)~and~(c)~) with GDM simulations (~(b)~and~(d)~) show a very good qualitative and quantitative agreement.
Having a look at the FW/BW ratios (Fig.~\ref{fig:fig4}, bottom row) we observe that uni-directional scattering occurs mostly towards the FW direction.
Spectrally relatively narrow peaks of strong forward scattering can be identified for both, TE and TM polarized illumination.
On the other hand, a zone of significant BW scattering is observed only in the case of TE polarization, similar to our observations on cylindrical SiNWs.

Selected spectra from Fig.~\ref{fig:fig4}, as well as several near-field intensity plots are shown in figure~\ref{fig:fig5}. 
The spectra in~(a-d) confirm the very good agreement between experiment (solid lines) and simulations (dashed lines). 
For the smallest nanowire widths as well as for long wavelengths, we find again an omnidirectional scattering, corresponding to a purely dipolar response (Fig.~\ref{fig:fig5}e). 
In analogy to cylindrical NWs, this first order resonance can only be excited in the TM geometry, TE polarized illumination results in a very weak scattering in the corresponding spectral range.
While small nanowires have spectrally broad optical resonances (figure~\ref{fig:fig5}a-b), narrower higher order resonances appear with increasing NW width (figure~\ref{fig:fig5}c-d). 
In contrast to the omnidirectional TM scattering in the smallest wires, the first occurrences of strong forward scattering seem to be the result of interference between a sharp quadrupolar resonance and a broad dipolar contribution.
We deduce this conclusion from the similarity in field intensity distributions in cylindrical and rectangular NWs (compare Figs.~\ref{fig:fig3}d and~\ref{fig:fig5}e,f,i).

Under TM excitation, we also observe branches of a different kind of Fano-like resonances where the FW/BW ratio is almost unity (see Fig.~\ref{fig:fig4}c-d and Fig.~\ref{fig:fig5}c-d, bottom row).
In a narrow spectral window the otherwise strong forward scattering is suddenly suppressed while BW scattering increases. 
Resonances with such field profiles are observed neither under TE polarization, nor in symmetric SiNWs ({\color{blue}see SI Figs.~S11-S15}, where scattering from rectangular wires is compared to different symmetric geometries).
We attribute these sharp features to horizontal guided modes along the SiNW width (\(X\)-direction), the wire side facets acting as the Fabry-Perot cavity mirrors.
To verify this assumption we assess the effective index for the supported guided mode(s) assuming a 1D waveguide of infinite extensions in \(X\)- and \(Y\)-direction.
For the case of TM polarized illumination at \(\lambda=580\,\)nm, the corresponding first order guided mode along the width of the NW has an effective index of \(n_{\text{eff}}=3.45\) (\(n_{\text{Si}}\approx 4.0\)). 
The first and second order standing wave patters in figure~\ref{fig:fig5}g and~h match perfectly with the nanowire widths of \(180\,\)nm, respectively \(330\,\)nm (at \(\lambda=560\,\)nm for the latter).
For more details on the supported guided modes, see {\color{blue}section~V in the SI}.
It is possible to exploit these Fabry-Perot like modes in order to toggle between FW and omni-directional scattering within a narrow spectral window -- simply by switching the incident polarization from TE to TM (see \textit{e.g.} around \(560\,\)nm in figure~\ref{fig:fig5}d).

Figs.~\ref{fig:fig5}(g-h) (TM: center row) correspond to near-field intensity distributions inside the SiNW at the first two orders of these TM-excited Fabry-Perot resonances. 
The field patterns inside the NW cross section correspond indeed to standing wave patterns of guided modes, which eventually lead to an almost equal FW and BW scattering in a narrow spectral region (see Fig.~\ref{fig:fig5}(c-d) bottom, \(550\,\text{nm} \lesssim \lambda \lesssim 600\,\text{nm}\)).
We remark that the assumption of spectrally sharp, guided modes between the NW side facets is also in agreement with former studies on scattering from rectangular dielectric NWs.\cite{fan_optical_2014, landreman_fabry-perot_2016}

\added{
Finally, we want to assess the influence of non-normal illumination on FW and BW scattering from rectangular SiNWs.
For spectroscopy on single nanowires, we need to use a high-NA microscope objective, which results in a large dispersion of incident angles, with angles up to \(\approx 72^{\circ}\) in our setups (NA\,\(0.95\), see also Methods).
In figure~\ref{fig:fig6}, we show simulated field intensity distributions in the scattering plane and scattering spectra for a rectangular SiNW of \(250\,\)nm width (height~\(90\,\)nm) for different incident angles (depicted in figure~\ref{fig:fig6}a). 
To calculate the spectra, while the angle of incidence is varied, we fix the integration solid angle to the upper (BW) and lower (FW) hemisphere, corresponding to a fixed position of the collecting microscope objective.
The wavelength in (b-g) is \(\lambda_0=600\,\)nm, the electric field is either polarized parallel to the scattering plane (TE) or along the NW axis (TM).
(b-g) show the field intensity and angular farfield intensity patterns in the NW section for increasing incident angles.
}
\added{
We observe, that while the original field distribution can still be recognized for angles up to \(\approx 36^{\circ}\), at high angles the nearfield distribution as well as the farfield radiation pattern are completely different.
In (h-j) we show FW scattering (h), BW scattering (i) and FW/BW ratio (j) spectra for the same rectangular SiNW. 
The black lines correspond to normal incidence and the green dashed lines to an incident angle of \(72^{\circ}\). 
The magenta lines denote the average scattering for illumination angles from \(0^{\circ}\) to \(72^{\circ}\), hence an approximation to our experimental conditions.
The spectra for an illumination angle corresponding to NA\,0.95 show often even completely opposite trends compared to normal incidence.
In (j) for instance regions exist where the directionality flips from FW under normal incidence to BW for an angle of \(72^{\circ}\).
On the other hand, the spectra calculated using the average illumination angles follow the trends of normal incidence. 
}
\added{
We conclude, that a normally incident plane wave gives a valid first-order approximation even for rectangular NWs illuminated by tightly focused beams.
This is also supported by the good agreement between normal-incident simulations and the experimental data (see figure~\ref{fig:fig4}).
% We note, that despite the general agreement between normal incidence simulations and dark-field spectra, some spectral features cannot be predicted using normally incident plane wave illumination. 
In the \textcolor{blue}{SI Figs.~S17-S19}, we show nearfield intensity maps for different excitation wavelengths and nanowire geometries as well as further spectra for comparison with figure~\ref{fig:fig5}, confirming the trends discussed above.
In the supporting informations, we also discuss a selected example, where the normal incidence approximation breaks down.
% More accurate modeling would require a rigorous description of the dark-field illumination,\cite{jiang_accurate_2015} which, however, is beyond the scope of the present work. 
}

%% ---------------------------------------------- SECTION: Conclusions
\section{Conclusions}

In conclusion, we demonstrated strongly directional light scattering from symmetric an asymmetric dielectric nanowires at optical frequencies.
We derived the conditions for exclusive forward or backward scattering in cylindrical nanowires from Mie theory and showed both theoretically and experimentally that anisotropic scattering can be obtained from individual silicon nanowires.
Unlike 0D-structures, directionality is a result of interference between different orders of either transverse electric or -magnetic modes.
Also, compared to 0D nano-particles the incident light polarization offers an additional degree of freedom in nanowires -- apart from material and geometry variations.
Besides being able to adjust the scattering intensity, the directionality and the spectral width of the resonances by varying the NW width, it is for example possible to switch between FW and BW scattering simply by rotating the polarization of the incident light.
Finally we showed that by breaking the symmetry of cylindrical nanowires and using rectangular cross sections instead, Fano-like resonances occur, leading to spectrally very sharp features with unity FW/BW scattering ratio, due to guided modes along the NW width. 
Again, this isotropic scattering can be triggered by switching the incident polarization between TE and TM.
\added{We demonstrated that asymmetric nanowires also allow to tune the scattering response via the angle of incidence.}
Our results open perspectives for various applications where \added{spectrally tunable} directional guiding of light is required at a nano scale\added{, for instance in optical nano-circuits}.
\added{Furthermore,} dielectric nanowires may be used for example for \replaced{anisotropic cloaking purposes, hence observer-dependent invisibility,}{cloaking purposes} in \added{wavelength-selective} field-enhanced spectroscopies \added{or for directional non-linear emission}.
\replaced{Finally, due to the simplicity of the silicon nanowires and their compatibility with production-ready technology, our results render nanowires very interesting for}{ or for} light harvesting applications in photovoltaics.

%% ---------------------------------------------- SECTION: Methods
\section{Methods}

\subsection{Crystal Growth of Cylindrical SiNWs}\label{sec:methods_vls_sinws}

The cylindrical silicon nanowires were grown in a hot-wall reduced pressure chemical vapor deposition system via the Vapor-Liquid-Solid (VLS) process.
Au droplets were used as catalysts, their sizes can be easily controlled and define the diameter of the nanowires. 
The reactor pressure was \(4.5\,\)Torr, as precursor for the Si growth silane (SiH\(_4\)) was used, diluted with hydrogen as carrier gas.
More details can be found in reference~\onlinecite{dhalluin_silicon_2010}.
Subsequent to the expitaxy, the grown SiNWs were removed from the substrate by careful sonication in isopropyl alcohol solution.
For the scattering experiments the NWs were finally drop-coated onto a transparent glass substrate with lithographic markers. 
A second sample was prepared on a silicon substrate for scanning electron microscopy, on which we determined a dispersion in NW diameters between around \(50\,\)nm and \(100\,\)nm.

%% ---------------------------------------------- SUBSECTION: Sample
\subsection{Lithographic SiNWs by Electron Beam Lithography \added{on Silicon on Quartz Substrates}}\label{sec:methods_ebl_sinws}

\added{The so-called silicon-on-quartz (SOQ) substrates were fabricated starting from commercially available SOI substrates. 
A first thinning step is realized by thermal oxidation and oxide removal in order to adjust the top silicon thicknesses according to specifications. 
Direct bonding is then performed to join the SOI to \replaced{fused silica}{quartz} substrates. 
As silicon and \replaced{silica}{quartz} present a large CTE mismatch (CTE: Coefficient of thermal expansion), high thermal processes are prohibited to improve the bonding strength and the annealing temperature is therefore limited to \(200^{\circ}\)C. \cite{moriceau_materials_2014}
This implies dedicated surface preparation to achieve high adherence energy at low temperature: SOI are prepared with a nitrogen plasma and the fused silica with a chemical mechanical process step. 
The bonded structure is then thinned by grinding and chemical etching in order to remove the base silicon substrate as well as the buried oxide layer (BOX).}

The rectangular NWs were fabricated by a top-down technique via electron beam lithography (EBL) and subsequent anisotropic plasma etching. \cite{han_realization_2011, guerfi_high_2013}
A transparent SOQ wafer with a \(90\,\)nm thick Si overlayer served as substrate.
A RAITH 150 writer at an energy of 30 keV was used for EBL, employing a thin (60nm) negative-tone resist layer (hydrogen silsesquioxane, ``HSQ''). 
After the EBL step, HSQ was developed by immersion in \(25\,\)\% tetramethylammonium hydroxide (TMAH) for \(1\,\)min. 
The patterns in the resist were finally written into the SOQ silicon top layer by reactive ion etching (RIE). 
For the RIE step, a SF\(_6\)/C\(_4\)F\(_8\) plasma was used. 
Etching was stopped when the \replaced{silica}{quartz} layer was reached, following in-situ analysis.
A residual layer of \(\approx 20\,\)nm of SiO\(_2\) (developed resist) remains on the structures, which has no major impact on the optical properties due to its low index and small thickness.
A comparison with a sample where the residual layer was removed using fluorhydric acid showed no significant difference in the spectra, but smaller structures were destroyed due to under-etching.

For SEM imaging, a second sample was prepared with identical process parameters on a silicon on insulator (SOI) substrate (Si-Layer: \(90\,\)nm, BOX layer: \(145\,\)nm)

\deleted{The SOQ substrates were fabricated starting from commercially available SOI substrates. 
A first thinning step is realized by thermal oxidation and oxide removal in order to adjust the top silicon thicknesses according to specifications. 
Direct bonding is then performed to join the SOI to quartz substrates. 
As silicon and quartz present a large CTE mismatch (CTE: Coefficient of thermal expansion), high thermal processes are prohibited to improve the bonding strength and the annealing temperature is therefore limited to \(200^{\circ}\)C. \cite{moriceau_materials_2014}
This implies dedicated surface preparation to achieve high adherence energy at low temperature: SOI are prepared with a nitrogen plasma and quartz with a chemical mechanical process step. 
The bonded structure is then thinned by grinding and chemical etching in order to remove the base silicon substrate as well as the BOX layer.}

%% ---------------------------------------------- SUBSECTION: Measurements
\subsection{Forward/Backward Confocal Darkfield Spectroscopy}\label{sec:methods_df_spectroscopy}
\paragraph*{Forward, ``FW'':} Transmission confocal optical dark-field (DF) microscopy was performed on an inverted optical microscope.
A broadband white lamp was focused through the sample substrate on the individual SiNWs using a DF condenser (Nikon, NA\,\(0.8-0.95\)).
The FW scattered light was collected using a Nikon NA\,\(0.75\) microscope objective, spatially selected via a confocal hole and guided through a polarization filter (\(\parallel\) or \(\perp\) to the NW axis for TM and TE geometry, respectively).
Finally, the light was dispersed by a \(150\,\)grooves per mm grating on a highly sensitive CCD (Andor Newton).

\paragraph*{Backward, ``BW'':} Reflection confocal optical DF experiments were carried out on a separate spectrometer (Horiba XploRA).
A white lamp was focused on individual SiNWs by a \(\times 50\) dark-field objective (NA\,\(0.5\), condenser: NA\,\(0.8-0.95\)).
The backscattered light was filtered by a confocal hole and a polarization filter and dispersed by a \(300\,\)grooves per mm grating on a highly sensitive CCD (Andor iDus 401).

In both setups (FW and BW), for practical reasons we put the polarization filters in the detection path (detection direction normal to the NW axis).
Because we limit our investigations to normal illumination (no depolarization) and due to the linearity of the fields, we assume that the polarization filter \textit{after} the sample is equivalent to a polarized, normal incidence.

\paragraph*{Normalization:} In the FW as well as in the BW scattering experiments, the intensity distribution of the respective lamp as well as the spectral response of the optical components was accounted for by the following normalization scheme:
First, the background noise was substracted (DF measurement on the bare SOQ / glass substrate).
Subsequently the data was normalized using the spectra of the respective lamps, measured through the bare substrate (``FW'' setup) and via a white reference sample (``BW'' setup).
Since we measure FW and BW scattering on two entirely different setups, we cannot compare the spectra quantitatively. 
Therefore we use the scattering from the smallest cylindrical nanowire (``NW1'', \(D\approx 50\,\)nm, Fig.~\ref{fig:fig3}a bottom) for normalization of the data between the two measurement geometries:
In sufficiently small NWs illuminated by TM polarized light, exclusively the fundamental dipolar TM\(_{01}\) mode is excited. 
This results in an omni-directional scattering, corresponding to a dipolar source along the SiNW axis.
Thus we can assume that the FW and BW scattered intensities under these conditions are of equal strength and normalize all spectra according to this reference.

\paragraph*{Incident and detection angles:} 
The upper limit for the incident angle from the dark-field condensers (FW and BW both \(\text{NA}=0.95\)) is
\begin{equation}
 \varphi_{\text{in}} = \sin^{-1}\left( \text{NA} \right) \approx 72^{\circ}.
\end{equation}
The high incident angles can modify the response of the Mie resonances. 
However, since the illumination is conic, only a fraction of the incident light has actually a large wavevector component parallel to the nanowire \(k_{\parallel}\). 
Thus, assuming normally incident plane wave illumination in the simulations is a valid approximation to our experimental conditions, further justified by the excellent agreement between our simulations and measurements.

The upper limits of the detection angles can be calculated by the collecting objectives' numerical apertures:
\begin{equation}
 \begin{aligned}
  \text{NA}_{\text{FW}} & = 0.75 
    \quad & \Rightarrow \quad 
  \varphi_{\text{FW}} &\approx 49^{\circ}\\
  \text{NA}_{\text{BW}} & = 0.5 
    \quad & \Rightarrow \quad 
  \varphi_{\text{BW}} &\approx 30^{\circ}.
 \end{aligned}
\end{equation}
With the maximum incident angle \(\varphi_{\text{in}} \approx 72^{\circ}\) this eventually leads to the following angular detection ranges (per polar quadrant):
\begin{equation}
 \begin{aligned}
  23^{\circ}\ & \ \lesssim \varphi_{\text{col., FW}} \lesssim \ & \ 120^{\circ} & \\
  42^{\circ}\ & \ \lesssim \varphi_{\text{col., BW}} \lesssim \ & \ 102^{\circ} &.
 \end{aligned}
\end{equation}
The simulated range is \(0^{\circ} \leq \varphi_{\text{col., sim.}} \leq 90^{\circ}\).
Given the excellent agreement between simulations and measurements, the deviations between  the collection angles seem not to cause a serious problem.
We attribute this to the fact that scattering is either omni-directional or else strongly forward/backward directional.
In both cases, large angles with respect to the incident wave-vector direction contribute only weakly to the total scattering.

See also {\color{blue}SI, Fig.~S16} for a detailed sketch, illustrating the collection angles.

\subsection{GDM Simulations}\label{sec:methods_gdm}

Because of the high aspect ratio of the nanowires, we assume infinitely long structures (NW axis along~\(Y\)). 
In this manner we simplify the computation to a two dimensional (2D) problem. 
The electric field is calculated using the Green dyadic method (GDM). 
Practically, we solve the vectorial Lippman Schwinger equation 
\begin{equation}\label{eq:LippmannSchwinger}
\begin{aligned}
 \mathbf{E}^{\text{2D}}(\mathbf{r}_{\parallel},\omega,\mathbf{k}_y)\ =\
 \mathbf{E}_0^{\text{2D}}(\mathbf{r}_{\parallel},\omega,\mathbf{k}_y)\ + \\ 
 \int \mathbf{G}_0^{\text{2D}}(\mathbf{r}_{\parallel},\mathbf{r}_{\parallel}', \omega, \mathbf{k}_y) \cdot 
 \chi(\omega) \mathbf{E}^{\text{2D}}(\mathbf{r}_{\parallel}',\omega,\mathbf{k}_y) d\mathbf{r}_{\parallel}'
\end{aligned}
\end{equation}
at the frequency~\(\omega\), by discretizing the nanostructure using square meshes. 
\(\mathbf{E}_0^{\text{2D}}\) and \(\mathbf{E}^{\text{2D}}\) are the incident field and resulting field inside the nanostructure, respectively. \(\mathbf{G}_0^{\text{2D}}\) is the Green's dyad and \(\chi\) the electric susceptibility of the structure material (\(\chi=\epsilon - 1\)).
\(\mathbf{k}_y\) is the incident wavevector along the NW axis. For normal incidence, \(\mathbf{k}_y=0\). 
The integral in Eq.~\eqref{eq:LippmannSchwinger} runs over the surface of the 2D particle cross section.
This self-consistent equation is solved (\textit{e.g.} using standard LU decomposition) and the field can be computed in the nanowire. 
Knowing the field inside the nanostructure, application of Eq.~\eqref{eq:LippmannSchwinger} permits to compute the field everywhere, both in the near-field and far-field zone. 
For details we refer the interested reader to Refs.~\onlinecite{martin_generalized_1995} and~\onlinecite{paulus_greens_2001}.

All simulations are carried out for SiNWs in vacuum (\(n_{\text{env.}}=1\)) to allow direct comparison with Mie theory for the cylindrical wires ({\color{blue}see SI, Figs.~S5-S10}).
Given the large index contrast with silicon (\(n_{\text{Si}}\approx 4\) in the visible), this is a good approximation, which is supported also by the excellent agreement between simulations and measurements.

%% ------- acknowledgements
%% ACHEMSO
% \begin{acknowledgement}

%% REVTEX
\begin{acknowledgments}

This work was supported by Programme Investissements d'Avenir under the program ANR-11-IDEX-0002-02, reference ANR-10-LABX-0037-NEXT, by LAAS-CNRS micro and nanotechnologies platform member of the French RENATECH network, by the R\'egion Midi-Pyr\'en\'ees and by the computing facility center CALMIP of the University of Toulouse under grant P12167.

%% REVTEX
\end{acknowledgments}

%% ACHEMSO
% \end{acknowledgement}

%% ------- SUP. INFO.

%% ACHEMSO
% \begin{suppinfo}

%% REVTEX
\section*{Supporting Informations}

In the supporting informations we show further details on FW/BW scattering from nanowires by Mie theory as well as additional simulations.

%% ACHEMSO
% \end{suppinfo}

\section*{Notes}

We note that a similar, theoretical work has been published during the review of this manuscript, describing scattering phenomena in nanowires with a radially anisotropic refractive index.\cite{liu_superscattering_2017}

The authors declare no competing financial interest.

% \bibliography{2017_FW_BW_SiNW}

\clearpage
\widetext

\section*{Supporting informations: \TITLE}
%
%
%%%%%%%%%% Merge with supporting info %%%%%%%%%%
%  \begin{center}
% \section*{Supporting Informations: EMO-GDM of double resonant silicon nanoantennas}
%  \end{center}
%%%%%%%%%% Merge with supplementary info %%%%%%%%%%
%%%%%%%%%% Prefix a "S" to all equations, figures, tables and reset the counter %%%%%%%%%%
\setcounter{equation}{0}
\setcounter{figure}{0}
\setcounter{table}{0}
\setcounter{page}{1}
\setcounter{section}{0}
\makeatletter
\renewcommand{\theequation}{S.\arabic{equation}}
\renewcommand{\thefigure}{S.\arabic{figure}}
\renewcommand{\bibnumfmt}[1]{[S#1]}
\renewcommand{\citenumfont}[1]{S#1}
%%%%%%%%%% Prefix a "S" to all equations, figures, tables and reset the counter %%%%%%%%%%

% \section{Cylindrical nanowires: Comparison Mie-theory vs. 2D-GDM simulations}

\section{Contributing Mie orders to FW/BW scattering intensity}

Mie theory can be applied to infinitely long cylinders by expanding the fields in vector cylindrical harmonics (instead of the vector spherical harmonics as in ``classical'' Mie theory for spherical particles). 
The expansion coefficients for the scattered fields are the so-called scattering coefficients \(a_n\) for electric multipole contributions and \(b_n\) magnetic multipoles.
Details can be found \emph{e.g.} in the book of Bohren and Huffman. \cite{bohren_absorption_1998}

% For infinitely long structures, cross sections have the unit of a length. We normalize the scattering and absorption cross sections by the projection of the geometrical section on the line perpendicular to the incidence (in our case the cylinder diameter). 
% In this way we obtain the scattering and absorption efficiencies \(Q_{\text{scat}}\) and \(Q_{\text{scat}}\).

In the next step, the asymptotic scattered field far from the structure has to be calculated. 
In Mie theory, the S-matrix, relating the scattered field \(\mathbf{E}_s\) with the incident field \(\mathbf{E}_i\), can be used for this means. 
Under normal incidence on an infinite cylinder, the S-matrix, connecting the incident and the asymptotic scattered field, writes
\begin{equation}\label{eq:scattering_via_Smatrix}
 \begin{bmatrix}
  E_{s, \text{TM}} \\
  E_{s, \text{TE}}
 \end{bmatrix}
  = 
e^{\mathrm{i} 3\pi/4} \, \sqrt{\dfrac{2}{\pi k R}} \, e^{\mathrm{i} k R}
 \begin{bmatrix}
  T_1 & 0    \\
  0   & T_2
 \end{bmatrix}
 \begin{bmatrix}
  E_{i, \text{TM}} \\
  E_{i, \text{TE}}
 \end{bmatrix}
\end{equation}
with the wavenumber \(k=2\pi/\lambda\) and \(R\) the distance to the cylinder origin. 
Under normal incidence the TM and TE polarized components of the scattered fields are proportional to the S-matrix components \(T_1\), respectively \(T_2\):
\begin{equation}\label{eq:SI_Smatrix}
 T_1 = b_0 + 2 \sum\limits_{n=1}^{\infty} b_n \cos(n \varphi) 
 \quad\quad\quad\quad
 T_2 = a_0 + 2 \sum\limits_{n=1}^{\infty} a_n \cos(n \varphi) 
\end{equation}
where \(\varphi\) is the polar angle of scattering in the cross-sectional plane of the cylinder. \(\varphi = 0^{\circ}\) corresponds to the forward scattering direction.

If two-dimensional objects of arbitrary cross section need to be calculated, Mie theory cannot be used and numerical simulations become necessary. 
The Green dyadic method\cite{martin_generalized_1995} (GDM) can be applied on infinitely long, two dimensional problems (``2D-GDM''), by deriving appropriate Green's tensors in a similar way to Mie theory for infinitely long cylinders. \cite{paulus_greens_2001}
In order to determine the scattering to the far-field, each discretization cell can be considered a dipolar emitter, whose emission to the far-field can be obtained using the two-dimensional Green's tensor. 
The coherent superposition of the dipolar emission from all discretization cells gives the total scattered field at any arbitrary location outside the discretized object.

The BW and FW scattered far-field intensities write for both, Mie theory and 2D-GDM simulations:
\begin{equation}\label{eq:SI_integrated_FW_BW_intensities}
 I_{s, \text{BW}} = \int\limits_{\pi/2}^{3\pi/2} |\mathbf{E}_{s}(\varphi)|^2 \text{d}\varphi 
 \quad\quad\quad\quad
 I_{s, \text{FW}} = \int\limits_{-\pi/2}^{\pi/2} |\mathbf{E}_{s}(\varphi)|^2 \text{d}\varphi.
\end{equation}

\begin{figure}[htb]
\centering
\includegraphics[width=\textwidth,page=1]{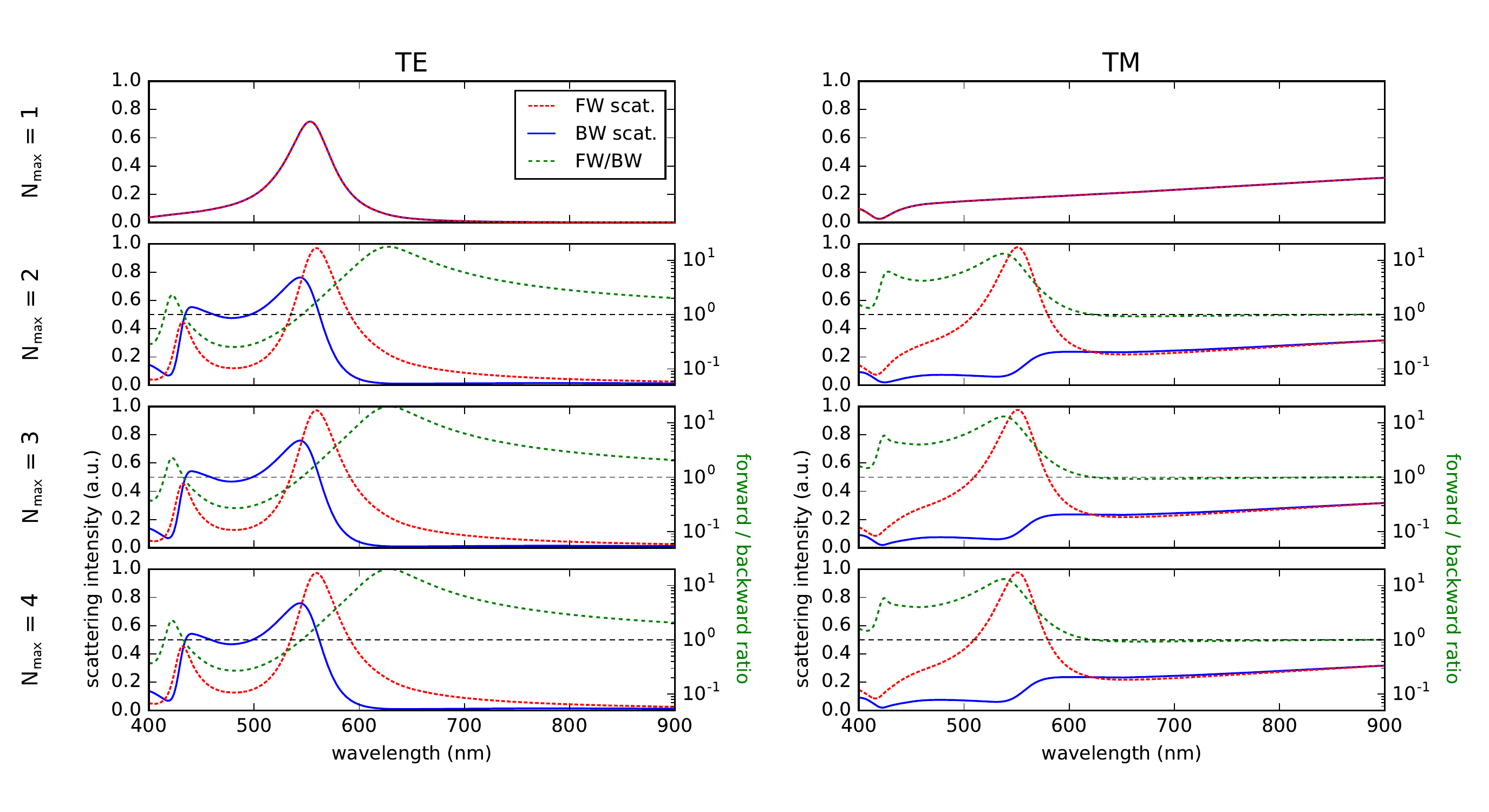}
\includegraphics[width=\textwidth,page=1]{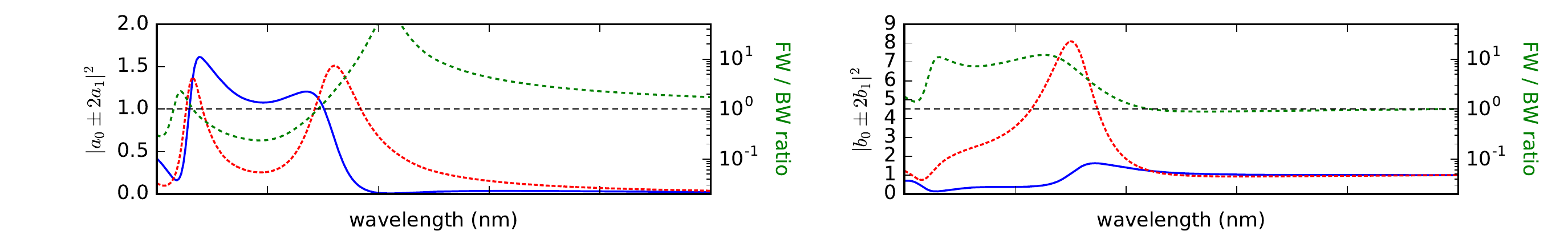}
\caption{\FIGCAPTIONPREFIX
Mie development of the FW/BW scattering from a SiNW with diameter \(D=100\,\)nm for the first 4 Mie coefficients \(a_n\) (TM: \(b_n\)). 
\(N_{\text{max}}\) corresponds to the number of Mie terms: \(n = 0\), \(n\in\{0,1\}\), \(n\in\{0,1,2\}\), \(n\in\{0,1,2,3\}\) (from top to bottom). 
Very bottom: FW/BW scattering intensity calculated using the Mie conditions derived for pure FW/BW scattering at \(\varphi=0^{\circ}\) / \(\varphi=180^{\circ}\) (no integration).
Left: TE, right: TM polarized normal incident plane wave. 
For better visibility, TE/TM data is normalized separately.}
\label{figSI:fig_N_max_order_D100nm}
\end{figure}

\begin{figure}[htb]
\centering
\includegraphics[width=\textwidth,page=1]{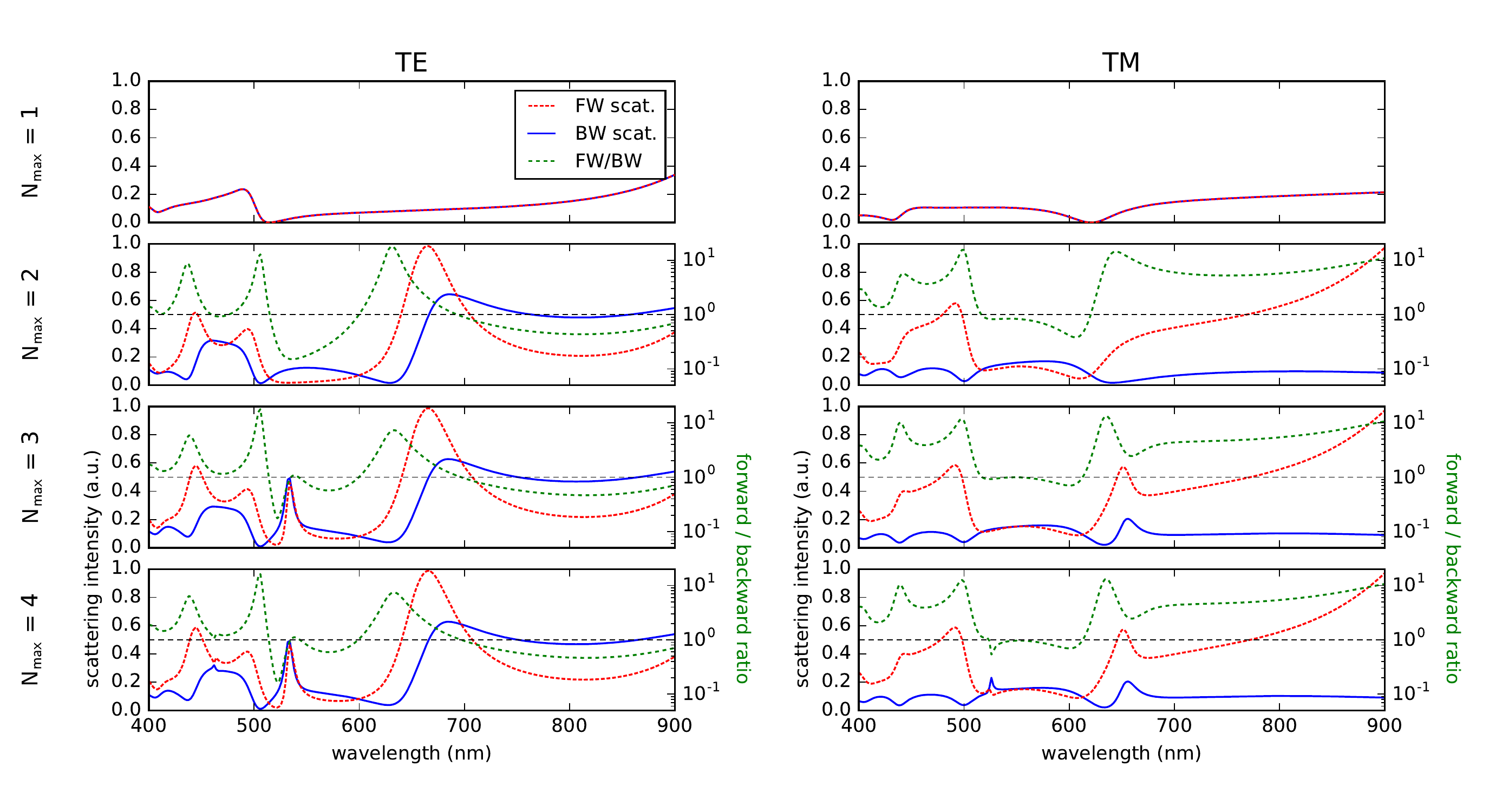}
\includegraphics[width=\textwidth,page=1]{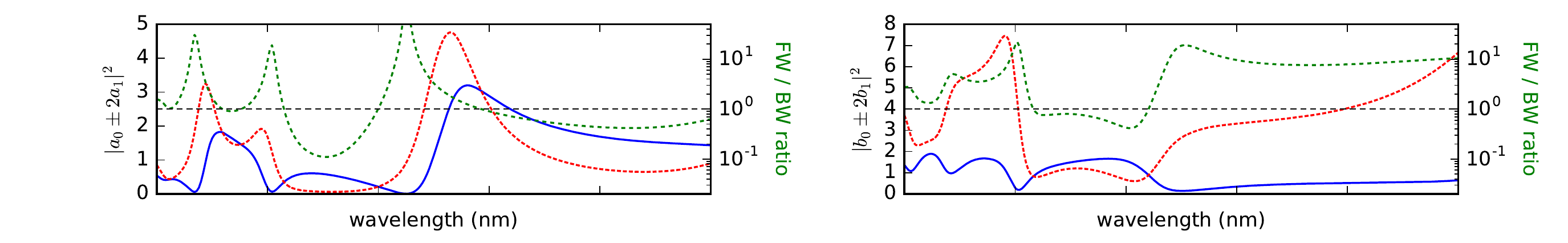}
\caption{\FIGCAPTIONPREFIX
Same as Fig.~\ref{figSI:fig_N_max_order_D100nm} for SiNW with diameter \(D=200\,\)nm.}
\label{figSI:fig_N_max_order_D200nm}
\end{figure}

Directional scattering is a result of interference between multiple simultaneously excited modes (see Eqs.~\eqref{eq:scattering_via_Smatrix}-\eqref{eq:SI_Smatrix}), and therefore the FW/BW resolved scattered intensity cannot be plotted individually for the different contributing scattering coefficients \(a_n\) and \(b_n\) (for TE and TM polarized normal incidence, respectively). 
In Figs.~\ref{figSI:fig_N_max_order_D100nm} and~\ref{figSI:fig_N_max_order_D200nm}, the FW and BW scattered intensity from a normally illuminated SiNW is calculated successively for an increasing number of contributing terms.
For a NW of \(D=100\,\)nm diameter (Fig.~\ref{figSI:fig_N_max_order_D100nm}), obviously only the first two orders contribute significantly, while for a larger NW (\(D=200\,\)nm, Fig.~\ref{figSI:fig_N_max_order_D200nm}), the interplay becomes more complex and several orders of the Mie scattering coefficients are necessary to describe all spectral features.
We want to stress the fact that despite some missing spectral features, using only the first two Mie orders gives already a very good approximation also in the case of larger SiNWs (see Fig.~\ref{figSI:fig_N_max_order_D200nm}).

We calculate furthermore the FW/BW scattering using the simple conditions derived in Eqs.~\eqref{eq:SI_FW_scat_firstorders_mie} and~\eqref{eq:SI_BW_scat_firstorders_mie}, \textit{i.e.} considering exclusively the FW (\(\varphi = 0^{\circ}\)) or BW (\(\varphi = 180^{\circ}\)) direction instead of performing the integration of Eq.~\eqref{eq:SI_integrated_FW_BW_intensities}.
The results are shown in the very bottom row of Figs.~\ref{figSI:fig_N_max_order_D100nm} and~\ref{figSI:fig_N_max_order_D200nm} and closely match the spectra obtained \textit{via} integration of a solid angle of \(\Delta\varphi = \pi\).

%%----------------------------------- FIGURE: cyl-SiNW Spectra ---
%%---------------------------------------------------------------
\enlargethispage{3cm}

%%----------------------------------- FIGURE: cyl-SiNW Spectra ---
%%---------------------------------------------------------------

\clearpage
\section{Mie scattering coefficients for an infinitely long cylinder}

\begin{figure}[htb]
\centering
\hspace{1cm}
\includegraphics[width=\textwidth,page=1]{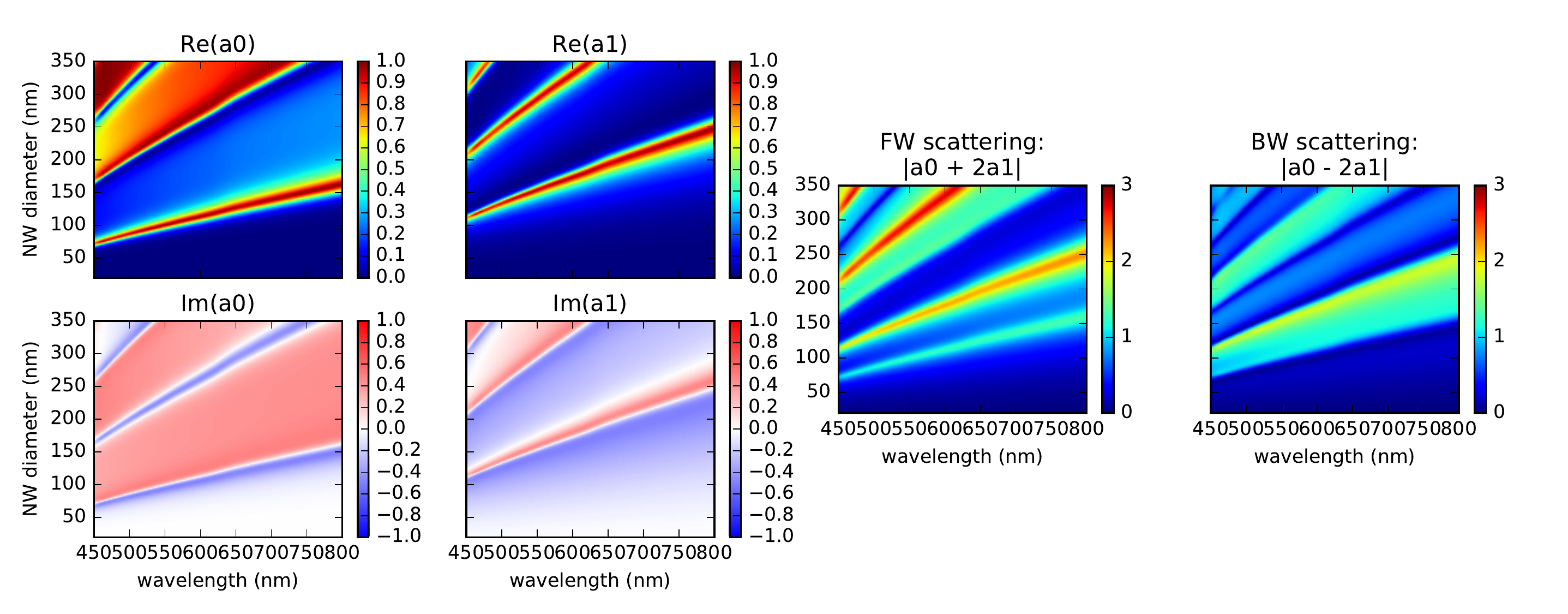}
\caption{\FIGCAPTIONPREFIX
Mie electric scattering coefficients \(a_0\) (left) and \(a_1\) (second left) as function of wavelength and cylinder diameter. Contributions to FW scattering (\(|a_0 + 2a_1|\), second right) and BW scattering (\(|a_0 - 2a_1|\), right) are shown separately. Top: real parts, bottom: imaginary parts.}
\label{figSI:Mie_coefficients_a0a1}
\end{figure}

\begin{figure}[htb]
\centering
\hspace{1cm}
\includegraphics[width=\textwidth,page=1]{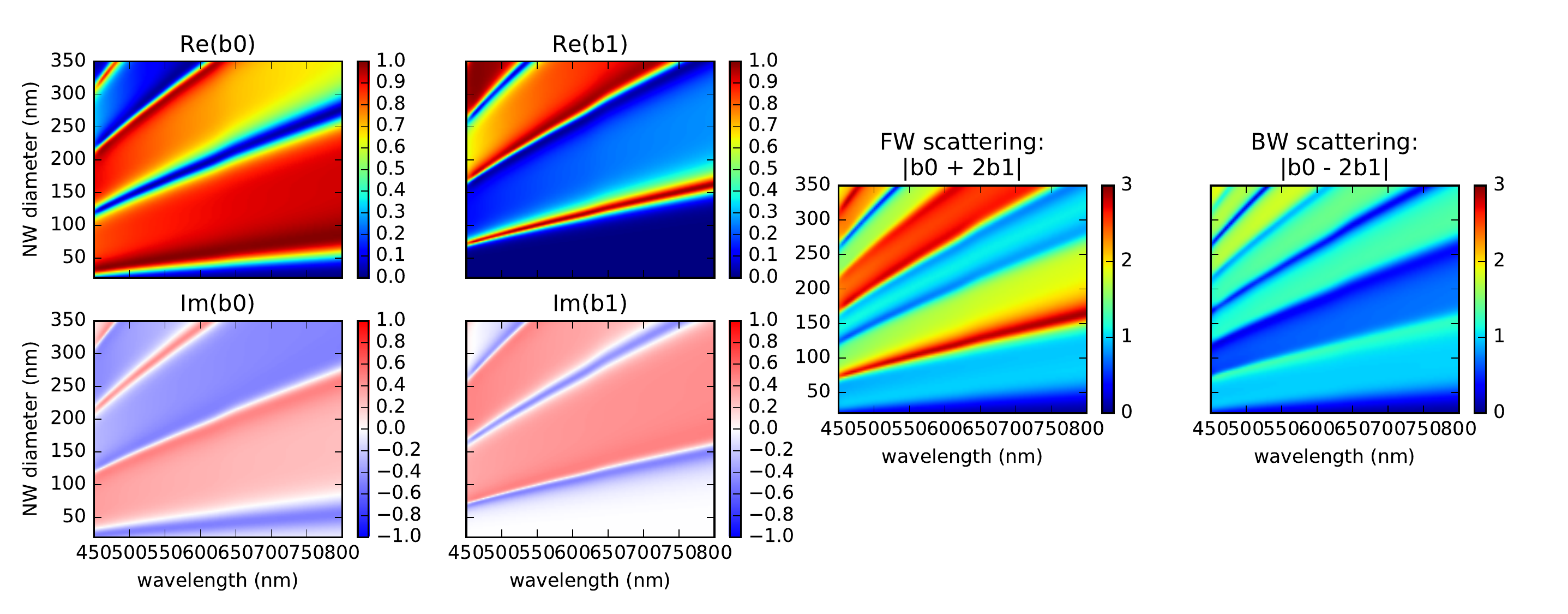}
\caption{\FIGCAPTIONPREFIX
Same as Fig.~\ref{figSI:Mie_coefficients_a0a1} but for the magnetic scattering coefficients~\(b_0\) and~\(b_1\).
% Mie magnetic scattering coefficients \(b_0\) (left) and \(b_1\) (second left) as function of wavelength and cylinder diameter. Contributions to FW scattering (\(b_0 + 2b_1\), second right) and BW scattering (\(b_0 - 2b_1\), right) are shown separately. Top: real parts, bottom: imaginary parts.
}
\label{figSI:Mie_coefficients_b0b1}
\end{figure}

The Mie scattering coefficients \(a_i\) and \(b_i\) (see \emph{e.g.} Ref.~\onlinecite{bohren_absorption_1998}) can be regarded as corresponding electric and magnetic multipole moments, representing the response of the nanowire to an external illumination.
As can be seen from Eqs.~\eqref{eq:scattering_via_Smatrix} and~\eqref{eq:SI_Smatrix}, under normal incidence on an infinitely long cylinder a TE polarized plane wave scattering occurs only due to electric multipole contributions (\(a_i\)) and a TM polarized excitation induces a purely magnetic scattering (\textit{via} \(b_i\)).
Furthermore, if considering only contributions of first and second order, the scattered fields in direct forward direction (\(\varphi=0^{\circ}\)) are for TM and TE polarization, respectively:
\begin{equation}\label{eq:SI_FW_scat_firstorders_mie}
 E_{s,\text{TM,FW}} \propto b_0 + 2 b_1
\quad\quad\quad\quad
 E_{s,\text{TE,FW}} \propto a_0 + 2 a_1
\end{equation}
while for the exact backward direction (\(\varphi=180^{\circ}\)) one finds:
\begin{equation}\label{eq:SI_BW_scat_firstorders_mie}
 E_{s,\text{TM,BW}} \propto b_0 - 2 b_1,
\quad\quad\quad\quad
 E_{s,\text{TE,BW}} \propto a_0 - 2 a_1.
\end{equation}
The first two order Mie coefficient \(a_n\) and \(b_n\) are shown in figures~\ref{figSI:Mie_coefficients_a0a1} and \ref{figSI:Mie_coefficients_b0b1}, respectively, separately for their real and imaginary parts. 
The absolute values of the sums/differences corresponding to pure FW/BW scattering (Eqs.~\eqref{eq:SI_FW_scat_firstorders_mie} and~\eqref{eq:SI_BW_scat_firstorders_mie}) are plotted as well (on the right).

%%--------------------------- FIGURE: spectra Mie coefficients a0,a1 ---
%%----------------------------------------------------------------------
\enlargethispage{3cm}

%%--------------------------- FIGURE: spectra Mie coefficients b0,b1 ---
%%----------------------------------------------------------------------

\clearpage
\section{Cylindrical nanowires: Comparison Mie-theory vs. 2D-GDM simulations vs. FDTD simulations}

\begin{figure}[htb]
\centering
\hspace{1cm}
\includegraphics[width=\textwidth,page=1]{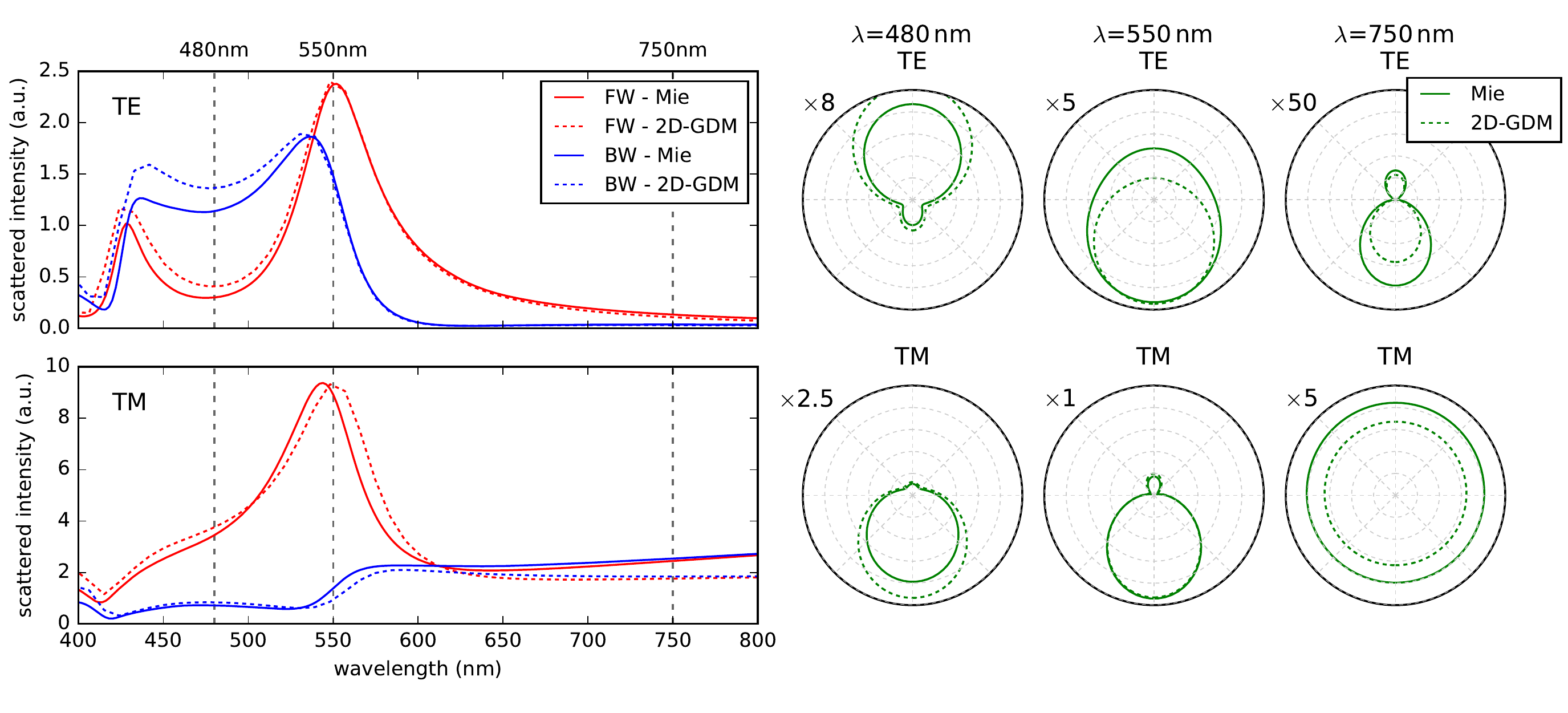}
\caption{\FIGCAPTIONPREFIX
Comparison between Mie theory (solid lines) and 2D-GDM simulations (dashed lines) of FW/BW scattered intensities as function of the wavelength of the incident plane wave, polarized TE (top) or TM (bottom). 
Right: Comparison of Mie/2D-GDM calculated radiation patterns at selected wavelengths, all normalized to the overall maximum (TM, \(\lambda=550\,\)nm. Normalization factors are shown on the upper right of each polar plot). 
All data for a SiNW of diameter \(D=100\,\)nm.}
\label{figSI:fig_compare_mie_2dgdm_spec_radpattern}
\end{figure}

Figure~\ref{figSI:fig_compare_mie_2dgdm_spec_radpattern} shows FW and BW scattering spectra for a \(D=100\,\)nm large SiNW, calculated by Mie theory (solid lines) and 2D-GDM simulations (dashed lines). 
Both, the qualitative and quantitative agreement is excellent, apart from a small spectral shift in the TM case and small intensity variations towards small diameters.

\begin{figure}[htb]
\centering
\hspace{1cm}
\includegraphics[width=\textwidth,page=1]{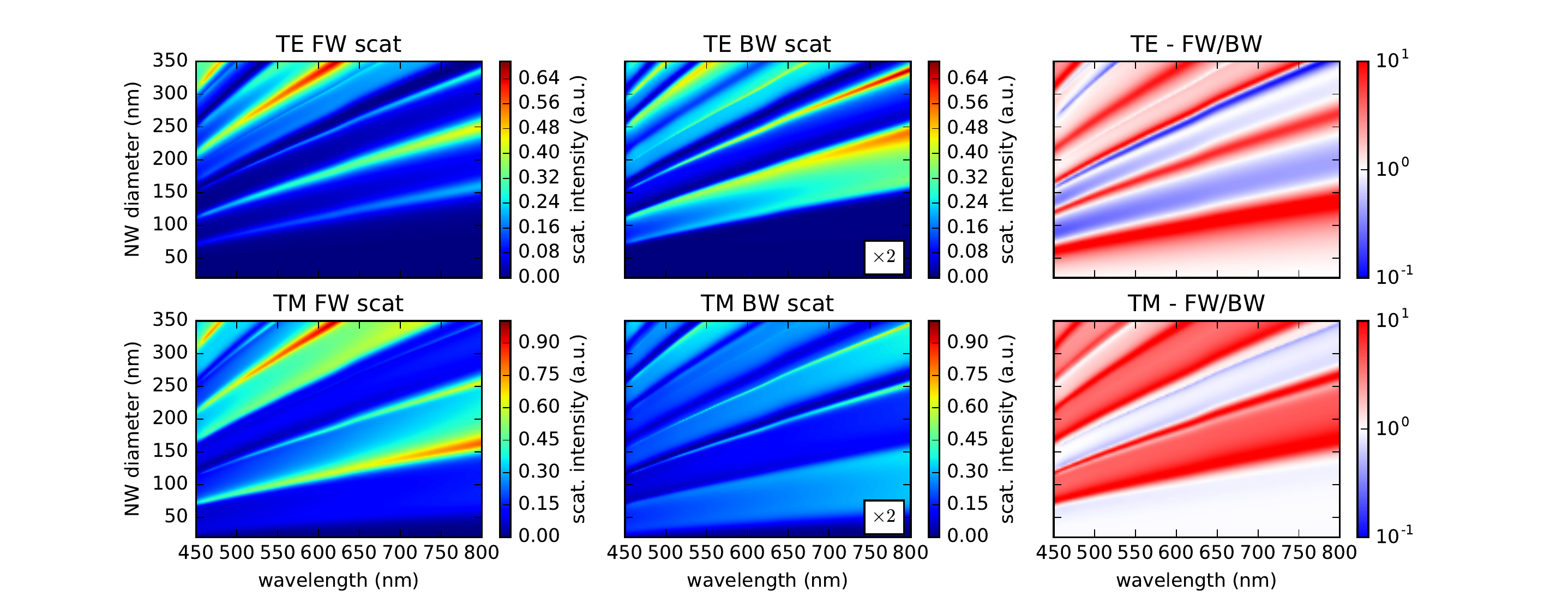}
\caption{\FIGCAPTIONPREFIX
FW/BW scattering intensities as function of NW diameter for an infinitely long silicon nanowire, calculated by Mie theory.}
\label{figSI:fig_compare_mie_2dgdm_cylnw_Mie}
\end{figure}

\begin{figure}[htb]
\centering
\hspace{1cm}
\includegraphics[width=\textwidth,page=1]{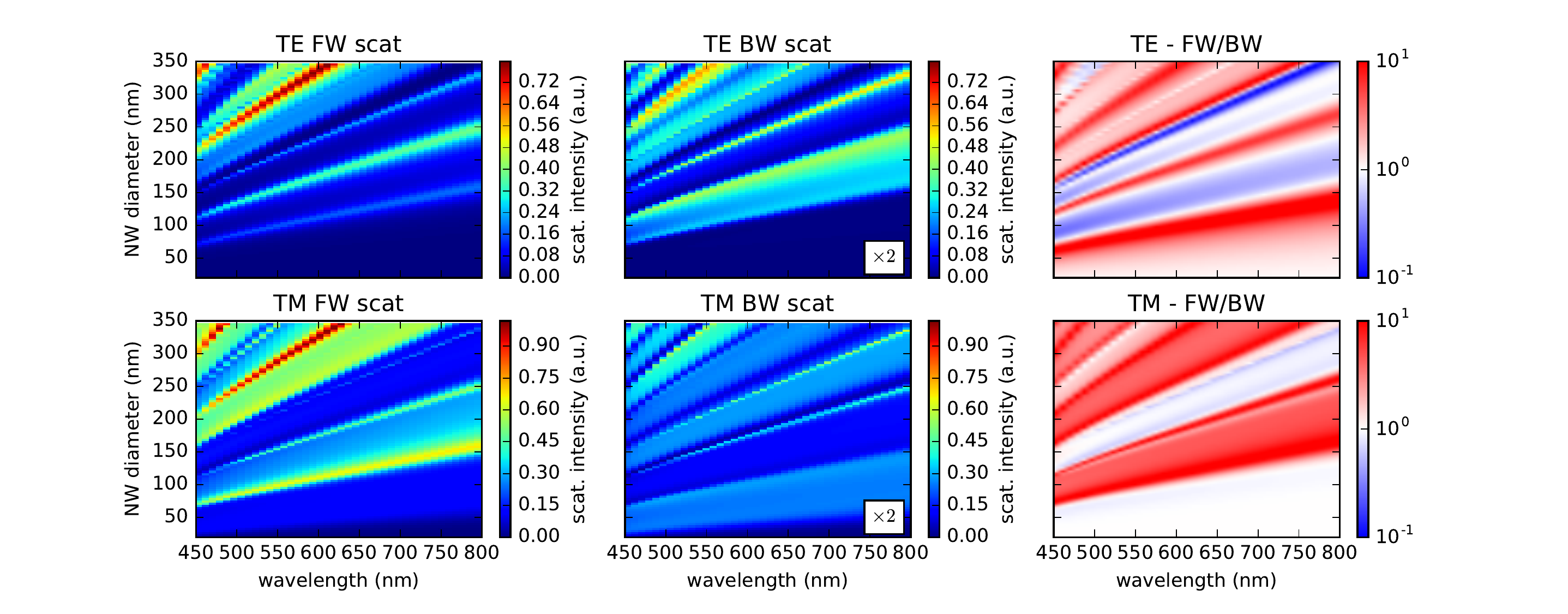}
\caption{\FIGCAPTIONPREFIX
Same as Fig.~\ref{figSI:fig_compare_mie_2dgdm_cylnw_Mie}, calculated using 2D-GDM.}
\label{figSI:fig_compare_mie_2dgdm_cylnw_GDM}
\end{figure}

Figures~\ref{figSI:fig_compare_mie_2dgdm_cylnw_Mie} and~\ref{figSI:fig_compare_mie_2dgdm_cylnw_GDM} further demonstrate the excellent agreement between analytical theory and 2D-GDM simulations, which justifies the use of the 2D-GDM for the comparison with our experimental results. 
This is particularly important in the case of our rectangular nanowires, which cannot be treated analytically.

%%-------------------------------- FIGURE: cyl-SiNW d100nm Spectra ---
%%-------------------------------------------------------------------

In a final step we double-check our methods by comparing them to finite difference time domain (FDTD) simulations. 
Therefore we calculate the electric and magnetic field intensity distributions inside infinitely long SiNWs using Mie theory, 2D-GDM simulation and \textit{via} the FDTD simulation toolkit ``Meep''\cite{oskooi_meep:_2010}. 
For the FDTD simulations we use the model from Ref.~\onlinecite{deinega_minimizing_2011} for the dispersion of silicon, for Mie and 2D-GDM we use tabulated data from Ref.~\onlinecite{palik_silicon_1997}.

\begin{figure}[htb]
\centering
\hspace{1cm}
\includegraphics[width=.9\textwidth,page=1]{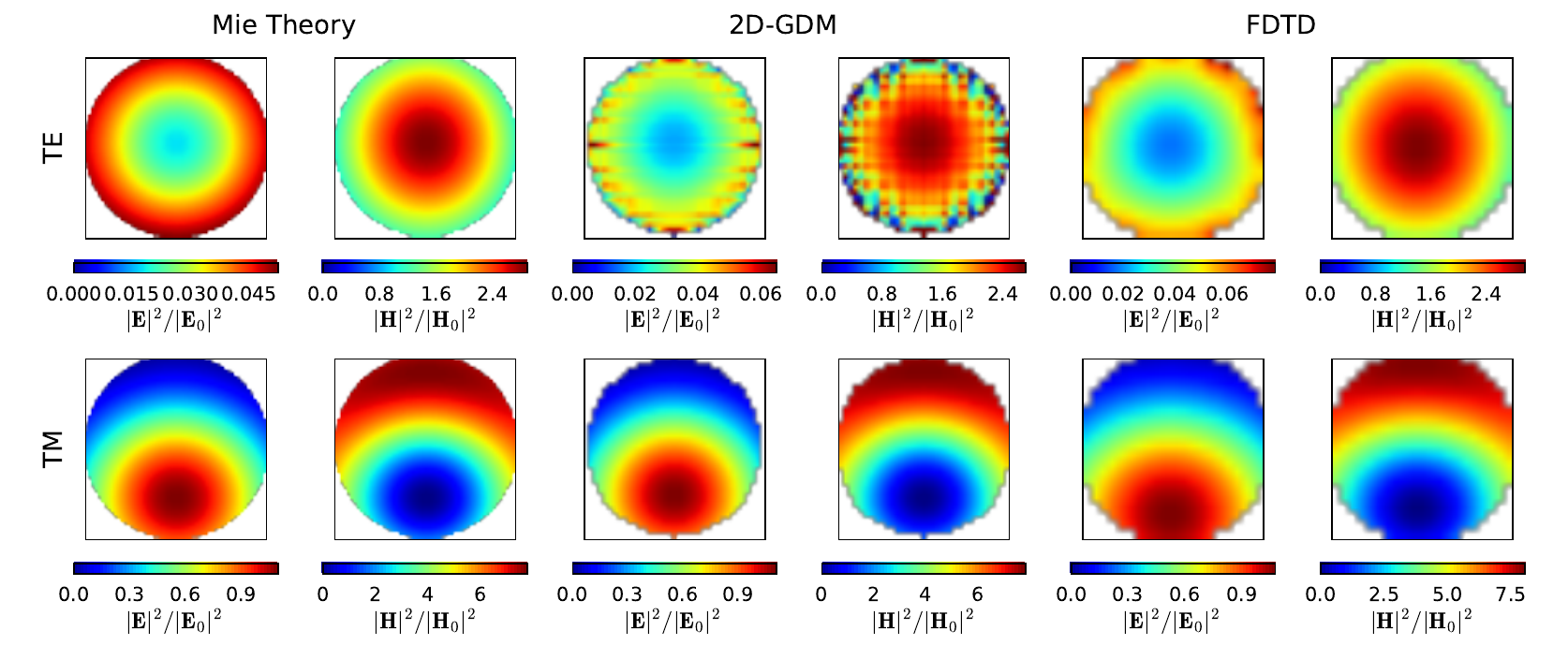}
\caption{\FIGCAPTIONPREFIX
Comparison of electric and magnetic field intensity distributions inside silicon nanowires, calculated with identical conditions using Mie theory (left), 2D-GDM simulations (center) and FDTD simulations (right). Incident plane wave with \(\lambda=500\,\)nm from the top, polarized normal to (TE, top row) or along (TM, bottom row) the NW axis. 
SiNW diameter is \(D=50\,\)nm.}
\label{figSI:compare_mie_2dgdm_meep_fields_d50nm}
\end{figure}

\begin{figure}[htb]
\centering
\hspace{1cm}
\includegraphics[width=.9\textwidth,page=1]{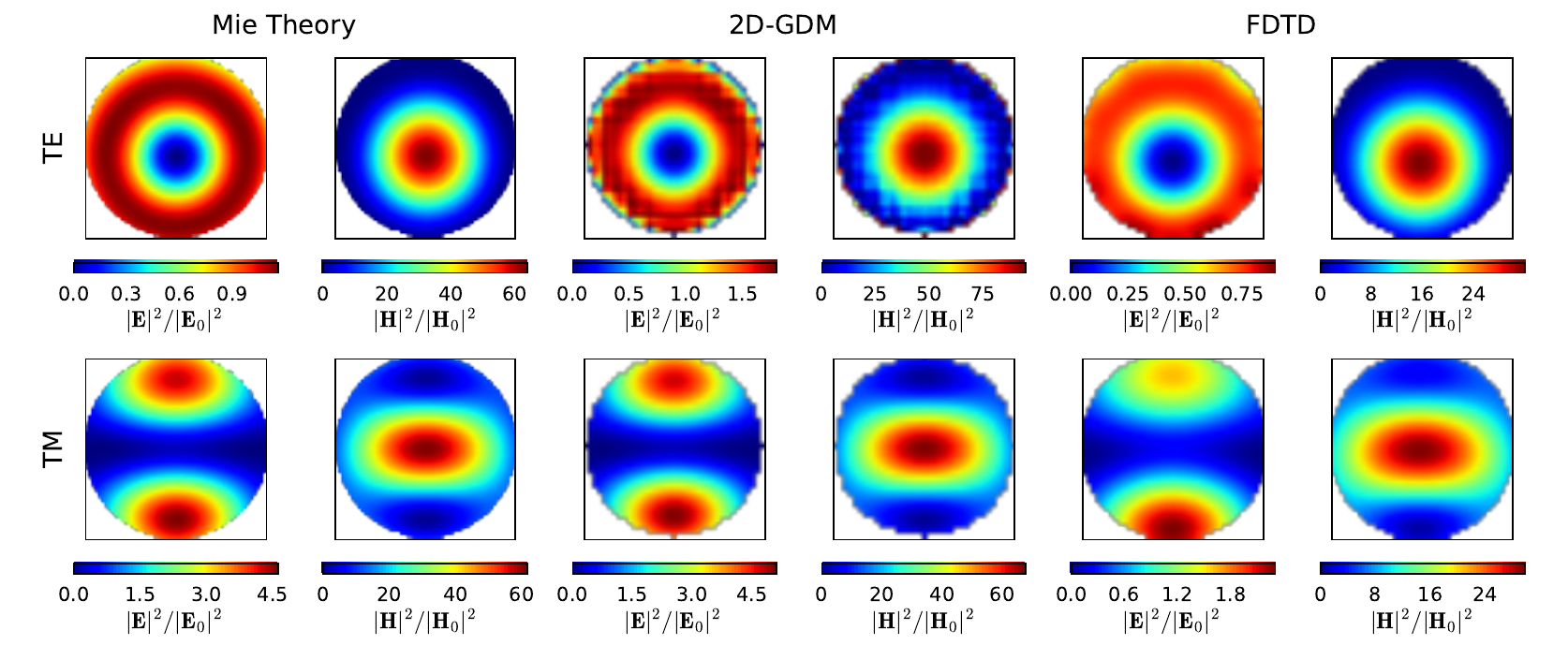}
\caption{\FIGCAPTIONPREFIX
Same as Fig.~\ref{figSI:compare_mie_2dgdm_meep_fields_d50nm}, with SiNW diameter of \(D=90\,\)nm.}
\label{figSI:compare_mie_2dgdm_meep_fields_d90nm}
\end{figure}

\begin{figure}[htb]
\centering
\hspace{1cm}
\includegraphics[width=.9\textwidth,page=1]{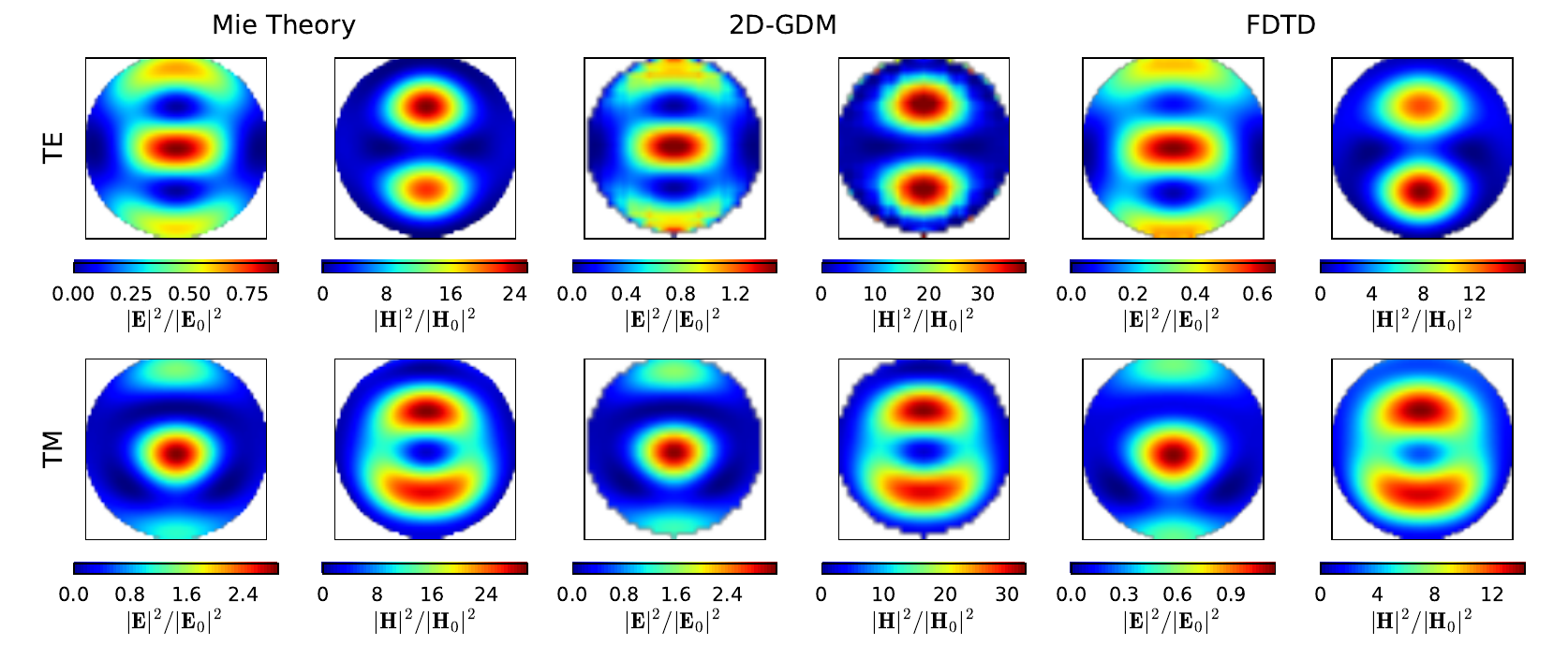}
\caption{\FIGCAPTIONPREFIX
Same as Fig.~\ref{figSI:compare_mie_2dgdm_meep_fields_d50nm}, with SiNW diameter of \(D=150\,\)nm.}
\label{figSI:compare_mie_2dgdm_meep_fields_d150nm}
\end{figure}

The results of this comparison are shown for selected SiNW diameters (\(D \in [50\,\text{nm},90\,\text{nm},150\,\text{nm}]\)) at fixed incident wavelength (\(\lambda=500\,\)nm) in figures~\ref{figSI:compare_mie_2dgdm_meep_fields_d50nm}-\ref{figSI:compare_mie_2dgdm_meep_fields_d150nm}. 
Again, we obtain a very good qualitative and quantitative agreement between the three methods.

We observe that 2D-GDM becomes more prone to numerical noise in the TE excitation configuration, which is probably due to the cubic mesh, not ideal in the description of a round structure such as a cylinder. 
In the TE configuration fields are not continuous along the air/NW interface, which is why in this case the numerical stability suffers more compared to a TM polarized incident plane wave. 
This might also explain the minor quantitative deviations between Mie and 2D-GDM, which are larger for TE than for TM polarization (see figure~\ref{figSI:fig_compare_mie_2dgdm_spec_radpattern}).

We note that FDTD can become quantitatively inaccurate for near field calculations when strong field confinement exists.\cite{forestiere_inverse_2016} 
This might explain the deviations in field intensities between FDTD and Mie/2D-GDM.

%%---------------------------------------------------------------
%% Colormaps FW/BW: Spectra Mie vs. GDM
%%---------------------------------------------------------------
\clearpage
%%----------------------------------- FIGURE: cyl-SiNW Spectra ---
%%---------------------------------------------------------------

%%----------------------------------- FIGURE: cyl-SiNW Spectra ---
%%---------------------------------------------------------------

%%---------------------------------------------------------------
%% E/B Fields inside NW
%%---------------------------------------------------------------
\clearpage
%%----------------------------- FIGURE: fieldplot comparison methods ---
%%-------------------------------------------------------------------

\enlargethispage{5cm}
%%----------------------------- FIGURE: fieldplot comparison methods ---
%%-------------------------------------------------------------------

%%----------------------------- FIGURE: fieldplot comparison methods ---
%%-------------------------------------------------------------------

\clearpage
\section{Comparison of FW/BW scattering from different nanowire geometries}

\begin{figure}[htb]
\centering
\hspace{1cm}
\includegraphics[width=.075\textwidth,page=1]{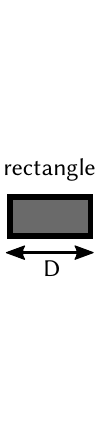}
\includegraphics[width=.85\textwidth,page=1]{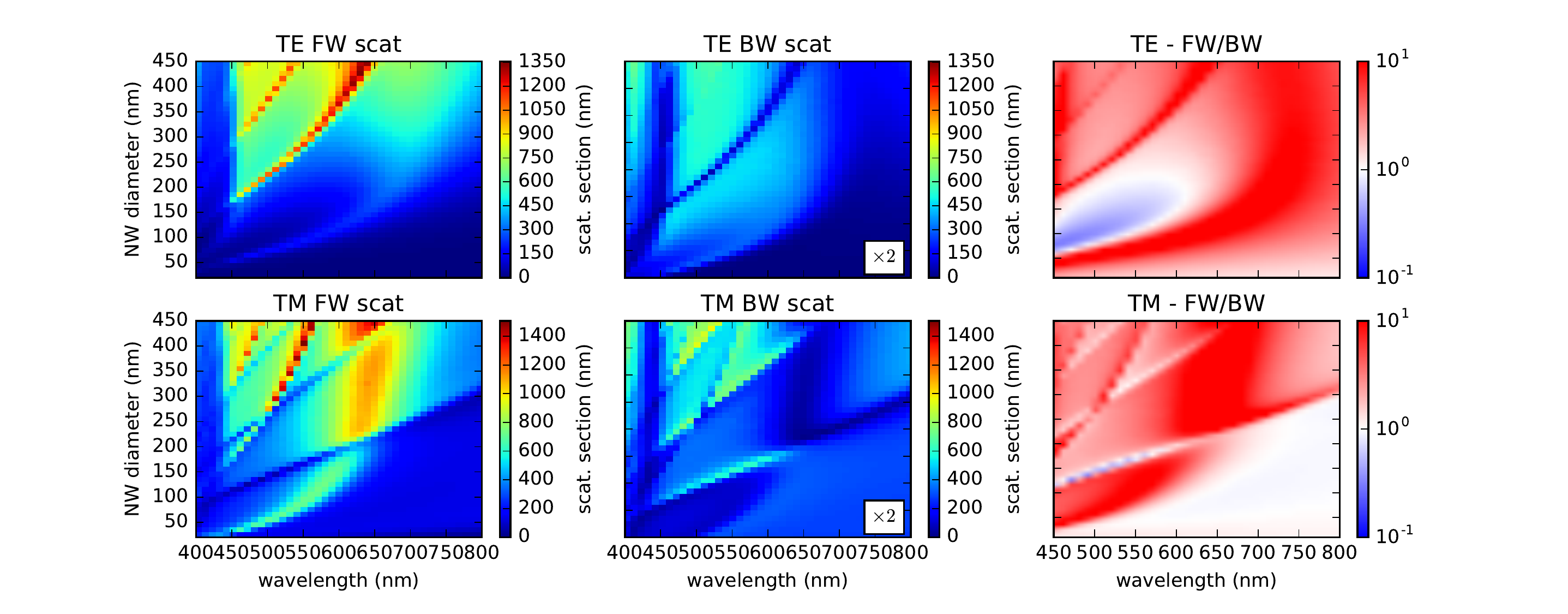}
\caption{\FIGCAPTIONPREFIX
FW/BW scattering from infinitely long SiNW of rectangular section (fixed height \(H=90\,\)nm).}
\label{figSI:fig_fwbwscat_geometry_rect}
\end{figure}

\begin{figure}[htb]
\centering
\hspace{1cm}
\includegraphics[width=.075\textwidth,page=1]{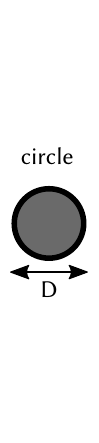}
\includegraphics[width=.85\textwidth,page=1]{"FW_BW_scat_NW_cylindrical"}
\caption{\FIGCAPTIONPREFIX
FW/BW scattering from infinitely long SiNW of cylindrical section.}
\label{figSI:fig_fwbwscat_geometry_cyl}
\end{figure}

\begin{figure}[htb]
\centering
\hspace{1cm}
\includegraphics[width=.075\textwidth,page=1]{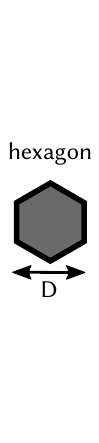}
\includegraphics[width=.85\textwidth,page=1]{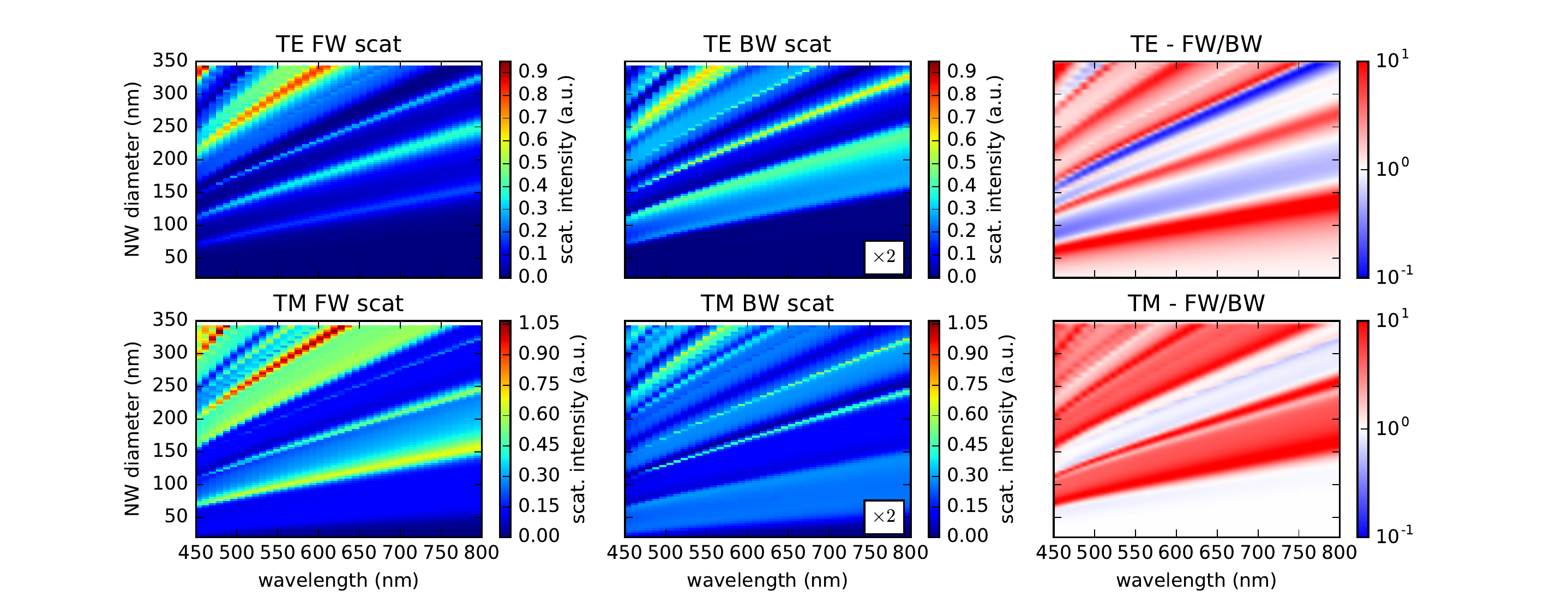}
\caption{\FIGCAPTIONPREFIX
FW/BW scattering from infinitely long SiNW of regular hexagonal section. The size parameter is the inscribed diameter of the hexagon.}
\label{figSI:fig_fwbwscat_geometry_hex}
\end{figure}

\begin{figure}[htb]
\centering
\hspace{1cm}
\includegraphics[width=.075\textwidth,page=1]{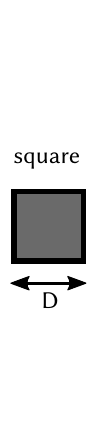}
\includegraphics[width=.85\textwidth,page=1]{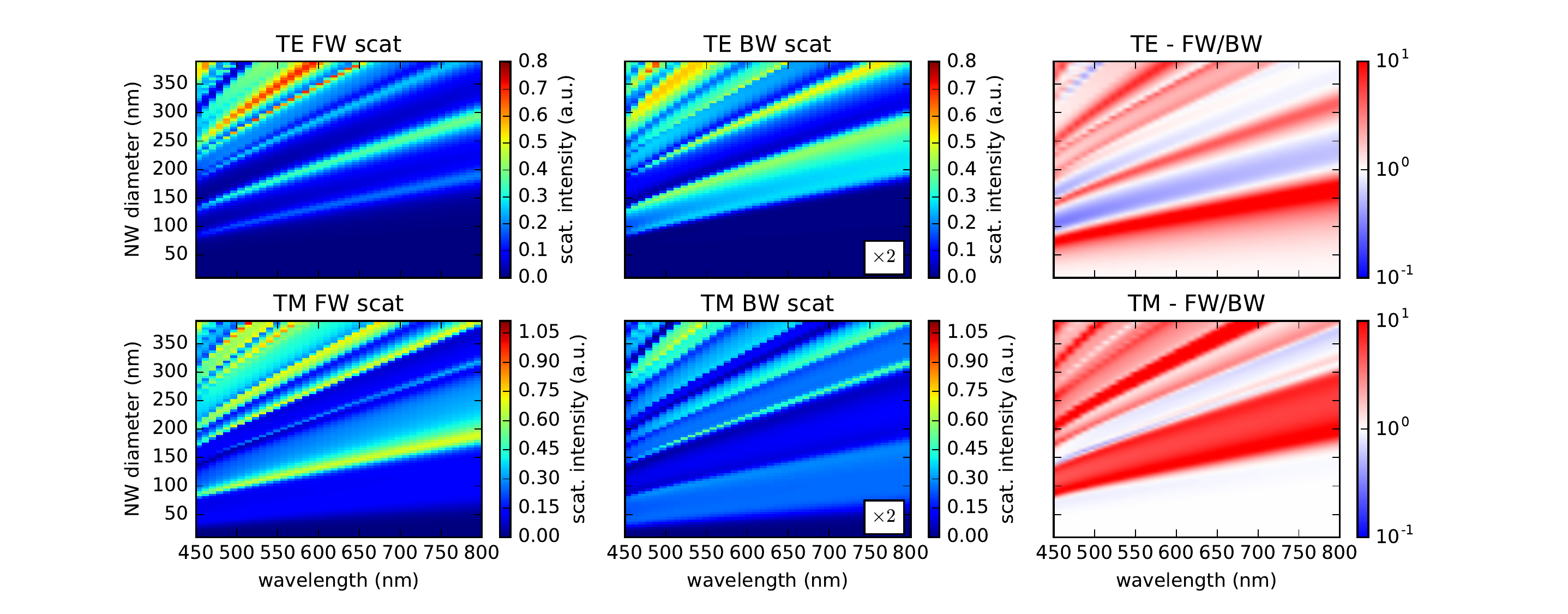}
\caption{\FIGCAPTIONPREFIX
FW/BW scattering from infinitely long SiNW of square section.}
\label{figSI:fig_fwbwscat_geometry_cube}
\end{figure}

\begin{figure}[htb]
\centering
\hspace{1cm}
\includegraphics[width=.075\textwidth,page=1]{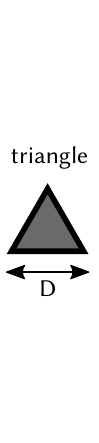}
\includegraphics[width=.85\textwidth,page=1]{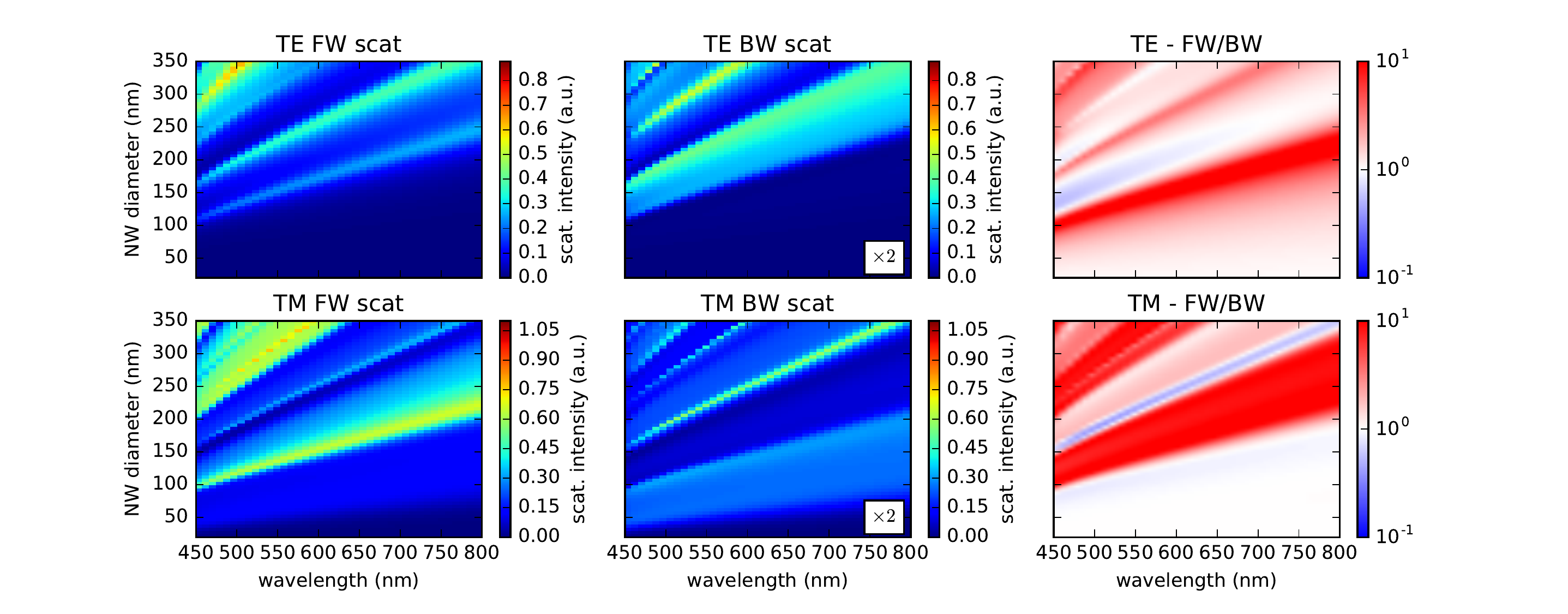}
\caption{\FIGCAPTIONPREFIX
FW/BW scattering from infinitely long SiNW of equilateral triangular section.}
\label{figSI:fig_fwbwscat_geometry_tri}
\end{figure}

We show below, that the Fano-like horizontally guided modes occur only in non-symmetric nanowires (\emph{e.g.} with rectangular cross sections, see case of TM polarization in Fig.~\ref{figSI:fig_fwbwscat_geometry_rect}).
Symmetric cross sections do not produce these kind of resonances (see Figs.~\ref{figSI:fig_fwbwscat_geometry_cyl}-\ref{figSI:fig_fwbwscat_geometry_tri})

%%----------------------------------- FIGURE: triangle-SiNW Spectra ---
%%---------------------------------------------------------------

%%----------------------------------- FIGURE: cyl-SiNW Spectra ---
%%---------------------------------------------------------------

%%----------------------------------- FIGURE: hex-SiNW Spectra ---
%%---------------------------------------------------------------

%%----------------------------------- FIGURE: cube-SiNW Spectra ---
%%---------------------------------------------------------------

%%----------------------------------- FIGURE: triangle-SiNW Spectra ---
%%---------------------------------------------------------------

\clearpage
\section{Laterally guided modes in rectangular nanowires of finit width}

It is possible to assess the effective index for the supported guided mode(s) assuming a 1D waveguide of infinite extensions in \(X\)- and \(Y\)-direction.\cite{hammer_1-d_nodate}
The thickness of this waveguide corresponds to the height of the rectangular nanowires (\(H=90\,\)nm).
To stay with the examples shown in the main paper figure~5, we calculate the supported modes and corresponding effective indices for the wavelengths \(\lambda=560\,\)nm and  \(\lambda=580\,\)nm (both: \(n_{\text{Si}}\approx 4.0\)).
Under TM incidence, ``TE'' guided modes are excited and the following fundamental guided modes are found:
\begin{itemize}
 \item \(580\,\)nm \(\quad\rightarrow\quad\) mode ``TE\(_0\)'': \(n_{\text{eff}}=3.45,\ \lambda_{\text{eff}}=170\,\)nm. \(\rightarrow\) first order standing wave pattern in the width of a NW of \(W=180\,\)nm.
 \item \(560\,\)nm \(\quad\rightarrow\quad\) mode ``TE\(_0\)'': \(n_{\text{eff}}=3.45,\ \lambda_{\text{eff}}=162\,\)nm. \(\rightarrow\) second order standing wave pattern in the width of a NW of \(W=330\,\)nm.
\end{itemize}

Under TE polarized incidence ``TM'' guided modes are induced. In this case, the effective index is much lower compared to the ``TE'' guided modes 
\begin{itemize}
 \item \(580\,\)nm \(\quad\rightarrow\quad\) mode ``TM\(_0\)'': \(n_{\text{eff}}=2.58,\ \lambda_{\text{eff}}=225\,\)nm.
\end{itemize}
In this case also the field confinement is significantly lower inside the slab, therefore the formation of standing wave patterns is less pronounced compared to the case of a TM polarized illumination of the nanowires.

We note that theoretically also higher order guided modes are allowed in the slab, however with very low effective indices and weak field confinements. 
Therefore the lowest order guided mode dominates and we never observed standing wave patterns of higher order guided modes.

% \clearpage
\section{Collection angles}

%%------------------------------ FIGURE: sketch collection angles ---
\begin{figure}[htb]
\centering
\includegraphics[width=\textwidth,page=1]{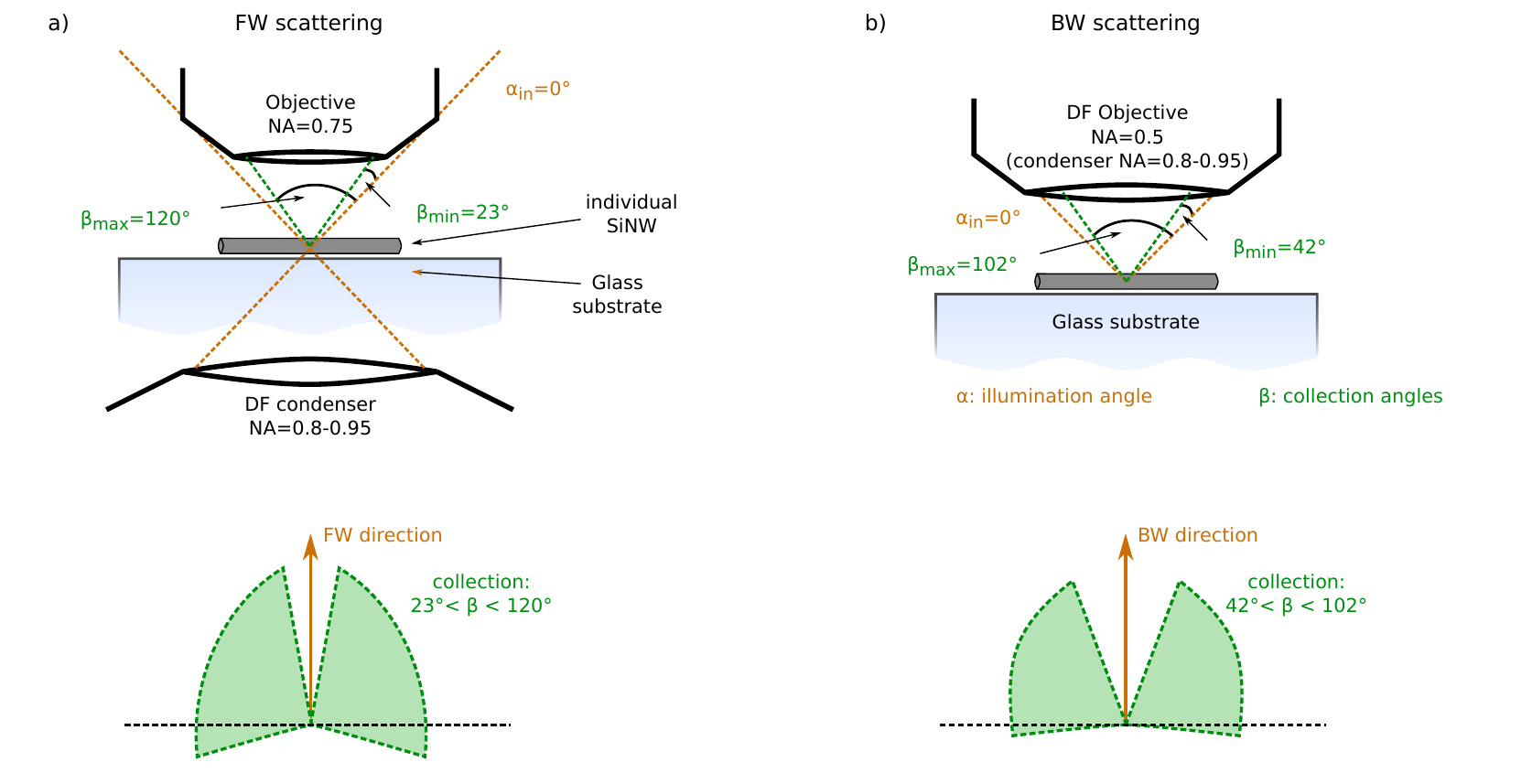}
\caption{\FIGCAPTIONPREFIX
Sketch illustrating the experimental collection angle range. (a) FW scattering setup, (b) BW scattering setup. Simulated angular range is \(0^{\circ} < \beta < 90^{\circ}\), corresponding to the entire upper, respectively lower hemisphere.}
\label{figSI:setup_angles}
\end{figure}
%%-------------------------------------------------------------------

\clearpage
\section{Oblique incidence}

\begin{figure}[htb]
\centering
\includegraphics[page=1]{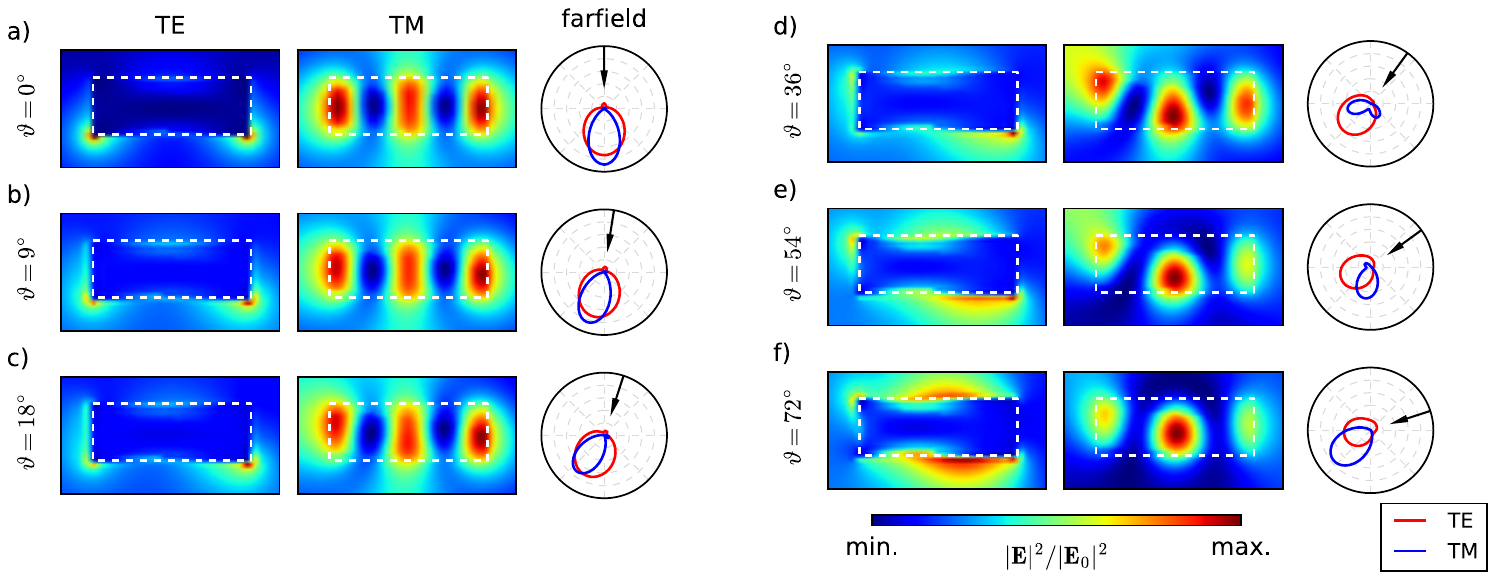}
\caption{\FIGCAPTIONPREFIX
Nearfield distributions in the sectional plane of a \(W=250\,\)nm, \(H=90\,\)nm rectangular NW under oblique incidence. 
Same as main paper, figure~6 (a-f), but for \(\lambda_0=705\,\)nm.}
\label{figSI:oblique_nearfields_w250}
\end{figure}

The fields in figure~\ref{figSI:oblique_nearfields_w250} are chosen for a spectral position, where for TM polarization the incident angle seems to have weak influence on the scattering (c.f. figure~6 (g-i) of main text).
By comparing the field patterns, it seems that a similar kind of guided mode is excited at this particular configuration for incidence either along \(Z\) (``top'') or along \(X\) (``side'').

%%----------------------- FIGURE: oblique inc. NF, W250nm, la705nm ---
%%-------------------------------------------------------------------

\begin{figure}[htb]
\centering
\includegraphics[width=\textwidth,page=1]{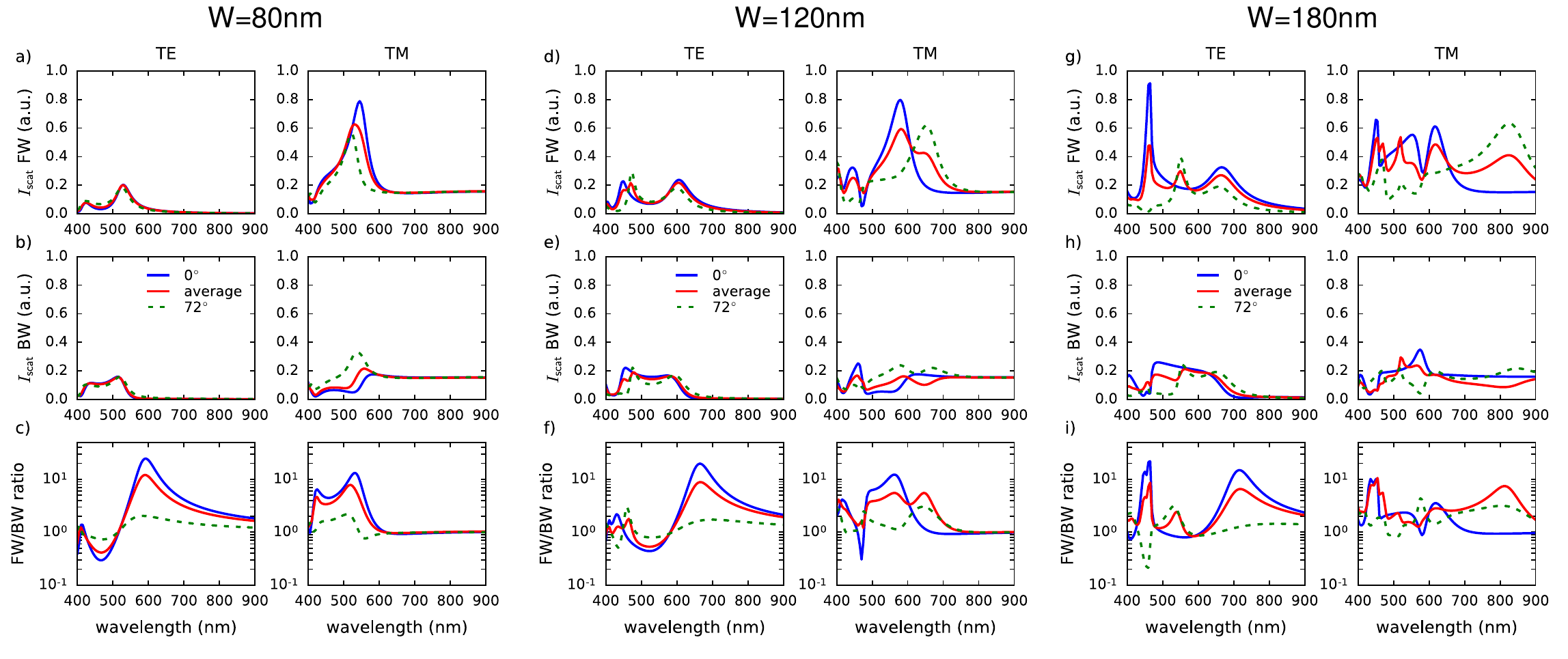}
\caption{\FIGCAPTIONPREFIX
FW/BW scattering spectra from rectangular NWs under oblique incidence. 
Same as main paper, figure~6 (g-i), for: (a-c) \(W=100\,\)nm, (d-f) \(W=150\,\)nm and (g-i) \(W=200\,\)nm.}
\label{figSI:oblique_spectra}
\end{figure}

Mostly, however, the angle of incidence does have a significant impact on the scattering from rectangular dielectric nanowires.
To assess this effect, we show in figure~\ref{figSI:oblique_spectra} additional spectra, analogously to main text figure~6, but for different NW widths.

%%------------------- FIGURE: oblique inc. spectra, W=100,150,200 ---
%%-------------------------------------------------------------------

For most of the spectra normal incidence gives a sufficient first order approximation compared to averaging over several incident angles.
However, for larger asymmetries and at long wavelengths, a FW-directed resonance occurs under oblique angles, which does not exist under normal incidence.

\begin{figure}[htb]
\centering
\includegraphics[page=1]{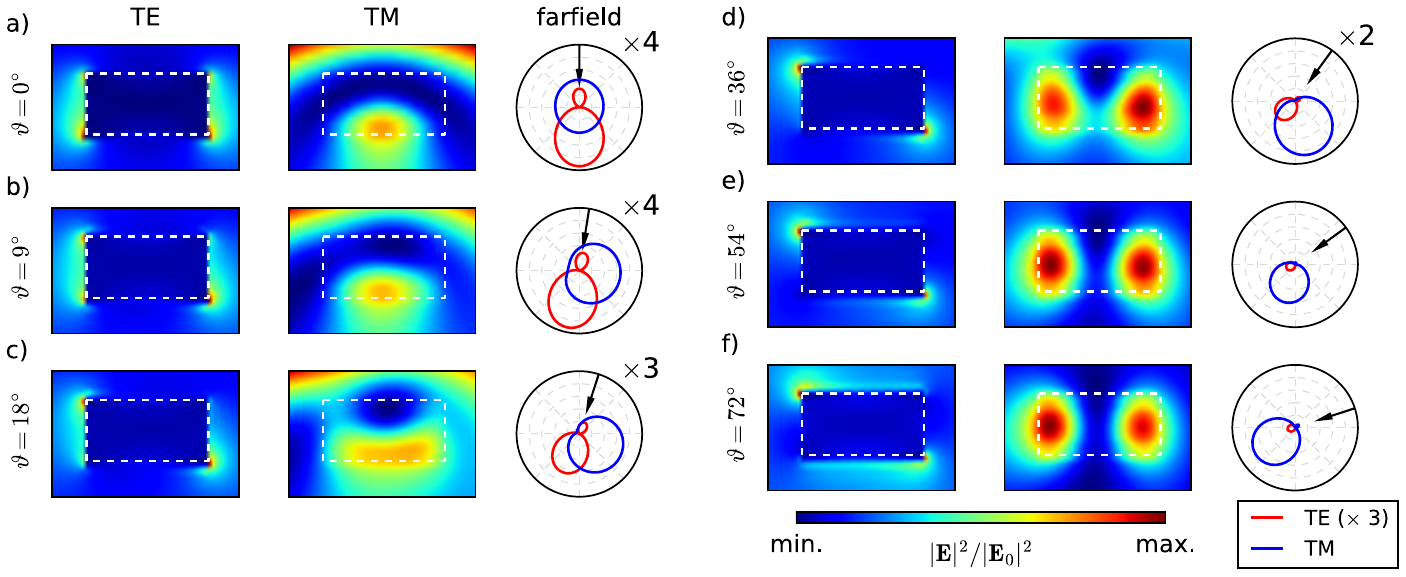}
\caption{\FIGCAPTIONPREFIX
Nearfield distributions in the sectional plane of a \(W=180\,\)nm, \(H=90\,\)nm rectangular NW under oblique incidence with \(\lambda_0=810\,\)nm.}
\label{figSI:oblique_nearfields_w180}
\end{figure}

In order to explain the remarkable deviation from normal incidence around the fundamental TM mode, we plot the nearfield distribution in the NW section for a \(W=180\,\)nm, \(H=90\,\)nm NW at \(\lambda_0=810\,\)nm (see figure~\ref{figSI:oblique_nearfields_w180}).
While under normal incidence, the fundamental mode along the ``height'' of the NW is excited (corresponding to a dipolar response), under oblique incidence the ``effective'' height of the NW increases rapidly due to the large aspect ratio.
Soon, a different resonance, corresponding to a significantly higher but narrower nanowire is excited, which contains a quadrupolar mode (see Fig.~\ref{figSI:oblique_nearfields_w180}d-f), leading to strongly directional scattering in that case.

%%----------------------- FIGURE: oblique inc. NF, W180nm, la810nm ---
%%-------------------------------------------------------------------

% \clearpage

\end{document}